\documentclass[aps,superscriptaddress,tightenlines,eqsecnum,nofootinbib]{revtex4}
\usepackage{graphicx,amssymb,amsbsy,bm,amsmath}
\usepackage{hyperref}
\newcommand{\be}{\begin{equation}}
\newcommand{\ee}{\end{equation}}
\newcommand{\bea}{\begin{eqnarray}}
\newcommand{\eea}{\end{eqnarray}}

\newcommand{\vp}{\ensuremath{\vec{p}}}

\begin{document}
\title{Constraints on dark matter
particles from theory, galaxy observations and N-body simulations.}
\author{D. Boyanovsky}
\email{boyan@pitt.edu} \affiliation{Department of Physics and
Astronomy, University of Pittsburgh, Pittsburgh, Pennsylvania 15260,
USA} \affiliation{Observatoire de Paris, LERMA. Laboratoire
Associ\'e au CNRS UMR 8112.
 \\61, Avenue de l'Observatoire, 75014 Paris, France.}
\affiliation{LPTHE, Universit\'e Pierre et Marie Curie (Paris VI) et
Denis Diderot (Paris VII), Laboratoire Associ\'e au CNRS UMR 7589,
Tour 24, 5\`eme. \'etage, 4, Place Jussieu, 75252 Paris, Cedex 05,
France}
\author{H. J. de Vega}
\email{devega@lpthe.jussieu.fr} \affiliation{LPTHE, Universit\'e
Pierre et Marie Curie (Paris VI) et Denis Diderot (Paris VII),
Laboratoire Associ\'e au CNRS UMR 7589, Tour 24, 5\`eme. \'etage, 4,
Place Jussieu, 75252 Paris, Cedex 05,
France}\affiliation{Observatoire de Paris, LERMA. Laboratoire
Associ\'e au CNRS UMR 8112.
 \\61, Avenue de l'Observatoire, 75014 Paris, France.}
\affiliation{Department of Physics and Astronomy, University of
Pittsburgh, Pittsburgh, Pennsylvania 15260, USA}
\author{N. G. Sanchez}
\email{Norma.Sanchez@obspm.fr} \affiliation{Observatoire de Paris,
LERMA. Laboratoire Associ\'e au CNRS UMR 8112.
 \\61, Avenue de l'Observatoire, 75014 Paris, France.}
\date{\today}
\begin{abstract}
Mass bounds on dark matter (DM) candidates are obtained for
particles that decouple in or out of equilibrium  while
ultrarelativistic with {\bf arbitrary} isotropic and homogeneous
distribution functions. A coarse grained Liouville invariant
primordial phase space density $ \mathcal D $ is introduced which
depends solely on the distribution function at decoupling. The
density  $ \mathcal D $ is explicitly computed and combined with
recent photometric and kinematic data on dwarf spheroidal satellite
galaxies in the Milky Way  (dShps) and the observed DM density today
yielding upper and lower bounds on the mass, primordial phase space
densities and velocity dispersion of the DM candidates. Combining
these constraints with recent results from $N$-body simulations
yield estimates for the mass of the DM particles in the range of a
few keV. We establish in this way a direct connection between the
microphysics of decoupling \emph{in or out} of equilibrium and the
constraints that the particles must fulfill to be suitable DM
candidates. If chemical freeze out occurs before thermal decoupling,
light bosonic particles can Bose-condense. We study  such
Bose-Einstein {\it condensate} (BEC) as a dark matter candidate. It
is shown that depending on the relation between the critical ($ T_c
$) and decoupling ($ T_d $) temperatures,  a BEC light relic could
act as CDM but the decoupling scale must be {\it higher} than the
electroweak scale. The condensate hastens the onset of the
non-relativistic regime and tightens the upper bound on the
particle's mass. A non-equilibrium scenario which describes particle
production and partial thermalization, sterile neutrinos produced
out of equilibrium and other DM models is analyzed in detail and the
respective bounds on mass, primordial phase space density and
velocity dispersion are obtained. Thermal relics with $ m \sim
\mathrm{few}\,\mathrm{keV} $  that decouple when ultrarelativistic
and sterile neutrinos produced resonantly or non-resonantly lead to
a primordial phase space density compatible with {\bf cored} dShps
and disfavor cusped satellites. Light Bose-condensed DM candidates
yield phase space densities consistent with {\bf cores} and if $
T_c\gg T_d $ also with cusps. Phase space density bounds on
particles that decoupled non-relativistically  combined with recent
results from N-body simulations suggest a potential tension for
WIMPs with $ m \sim 100\,\mathrm{GeV},T_d \sim \,10\,\mathrm{MeV} $.
\end{abstract}


\maketitle
\tableofcontents
\section{Introduction}

Although the existence of dark matter (DM) was inferred several
decades ago \cite{zwoo}, its nature still remains elusive. Candidate
dark matter particles are broadly characterized as cold, hot or warm
depending on their velocity dispersions. The clustering properties
of collisionless DM candidates in the linear regime depend on the
free streaming length, which roughly corresponds to the Jeans length
with the particle's velocity dispersion replacing the speed of sound
in the gas. Cold DM (CDM) candidates feature a small free streaming
length favoring a bottom-up hierarchical approach to structure
formation, smaller structures form first and mergers lead to
clustering on the larger scales.

Among the CDM  candidates are
weakly interacting massive particles (WIMPs) with $m \sim
10-10^{2}\,\mathrm{GeV}$. Hot DM (HDM) candidates feature large free
streaming lengths and favor top down structure formation, where
larger structures form first and fragment. HDM particle candidates
are deemed to have masses in the few $\mathrm{eV}$ range, and warm
DM (WDM) candidates are intermediate with  a typical mass range $ m
\sim 1-10 \,\mathrm{keV} $.

\medskip

The \emph{concordance} $ \Lambda\mathrm{CDM} $ standard cosmological
model emerging from CMB, large scale structure observations and
simulations favors the hypothesis that DM is composed of primordial
particles which are cold and collisionless \cite{primack}. However,
recent observations hint at possible   discrepancies with the
predictions of the $ \Lambda\textrm{CDM} $ concordance model: the
satellite and cuspy halo problems. \\

The satellite problem, stems from
the fact that CDM favors the presence of substructure: much of the
CDM is not smoothly distributed but is concentrated in small lumps,
in particular in dwarf galaxies for which there is scant observational
evidence so far. A low number of satellites have been observed in
Milky-Way sized galaxies \cite{kauff,moore,moore2,klyp}. This substructure is a
consequence   of the CDM power spectrum which favors small scales
becoming non-linear first, collapsing in the bottom-up
hierarchical manner and surviving the mergers as dense clumps
\cite{moore,klyp}. \\

The cuspy halo problem arises from the result of large scale
$N$-body simulations of CDM clustering which predict a monotonic
increase of the density towards the center of the halos
\cite{dubi,frenk,moore2,bullock,cusps}, for example the universal
Navarro-Frenk-White (NFW) profile $ \rho(r)\sim r^{-1}(r+r_0)^{-2}
$ \cite{frenk} which describes accurately clusters of galaxies, but
indicates a divergent cusp at the center of the halo. Recent
observations seem to indicate central cores in dwarf
galaxies \cite{dalcanton1,van,swat,gilmore}, leading to the 'cusps
vs cores' controversy. \\

A recent compilation of observations of dwarf spheroidal  galaxies
dSphs \cite{gilmore}, which are considered to be prime candidates for
DM subtructure \cite{spergel}, seem to \emph{favor} a core with a smoother central
density and a low
 mean mass density $ \sim 0.1\,M_{\odot}/\mathrm{pc}^3 $ rather
 than a cusp \cite{gilmore}. The data  cannot yet rule out cuspy
 density profiles which allow a maximum density $ \lesssim 60
\,M_{\odot}/\mathrm{pc}^3 $
  and the interpretation and analysis of the observations
 is not yet conclusive \cite{dalcanton1,van2}.  These \emph{possible} discrepancies
have rekindled an interest
 in WDM particles,  which feature a velocity dispersion larger
 than CDM particles, and consequently larger free-streaming lengths which smooth-out
the inner
 cores and would be prime candidates to relieve the cuspy
 halo and satellite  problems \cite{turok}.

\medskip

A possible WDM candidate is a sterile
 neutrino \cite{dw,este,kusenko} with a mass in the $\mathrm{keV}$ range and
produced via their mixing
 and oscillation with an active neutrino species either non-resonantly \cite{dw},
 or through MSW (Mikheiev-Smirnov-Wolfenstein) resonances in the medium \cite{este}. Sterile
 neutrinos can decay into a photon and an active neutrino (more
 precisely the largest mass eigenstate decays into the lowest one and a
 photon) \cite{pal} yielding the possibility of direct constraints on the
 mass and mixing angle from the diffuse X-ray background \cite{Xray}.

 \medskip

Observations of cosmological structure formation  via the Lyman-$\alpha$ forest
provide a complementary probe
 of primordial density fluctuations on small scales which yield an
 indirect constraint on the masses of WDM candidates. While
 constraints from the diffuse X-ray background yield an \emph{upper} bound
 on the mass of a putative sterile neutrino in the range
 $ 3-8\,\mathrm{keV} $ \cite{Xray}, the latest Lyman-$\alpha$
 analysis \cite{lyman}  yields \emph{lower} bounds in the range
 $ 10-13\,\mathrm{keV} $   in tension with the X-ray
 constraints.  More recent constraints from Lyman-$\alpha$ yield a lower limit for the
 mass of a WDM candidate $ m_{WDM} \gtrsim 1.2\,\mathrm{keV} \,(2\sigma) $ for an
early decoupled \emph{thermal}
 relic and $m_{WDM} \gtrsim 5.6\,\mathrm{keV} \,(2\sigma)$ for sterile neutrinos
\cite{viel}. Strong
 upper limits on the mass and mixing angles of sterile neutrinos have been recently
discussed \cite{beacom}, however,
 there are uncertainties as to whether WDM candidates can explain large cores in
dSphs \cite{strigari}.
  It has been recently argued \cite{palazzo} that if
 sterile neutrinos are produced non-resonantly \cite{dw} the combined
 X-ray and Lyman-$\alpha$ data suggest that  these \emph{cannot} be the \emph{only}
WDM component, with an upper limit for  their
 fractional relic abundance  $ \lesssim 0.7 $. Recent \cite{boyarski2} constraints on a
 radiatively decaying DM particle from the EPIC spectra of (M31) by XMM-Newton
confirms this result and  places
 a stronger lower mass limit $ m < 4\,\mathrm{keV} $.

All these results suggest  that DM could be a mixture of several components with
sterile neutrinos as viable candidates.

 \medskip

  {\bf Motivation and goals:} Although the $\Lambda\mathrm{CDM}$ paradigm describes
 large scale structure formation remarkably well, the \emph{possible} small scale
discrepancies mentioned above  motivate us to  study new constraints
that
 different dark matter components must fulfill to be suitable candidates.
  Cosmological bounds on  dark matter components
  primarily focused on standard model neutrinos \cite{bond,TG}, heavy relics that
decoupled in local thermodynamic  equilibrium (LTE) when
 non-relativistic \cite{LW,kt,dominik} or \emph{thermal ultrarelativistic
 relics} \cite{madsen,madsenbec,madsenQ,salu,hogan}. More recently, cosmological
precision data were used to constrain the (HDM) abundance of low mass particles
\cite{pastor, steen,raffelt,mena} assuming these to be thermal relics.

\medskip

The main results of this article are:

\medskip

(\textbf{a:}) We consider particles that  decouple {\bf in or  out of
 LTE} during the radiation dominated era with  an \emph{arbitrary}
(but homogeneous and isotropic) distribution  function.  Particles
which decouple being ultrarelativistic  eventually become
non-relativistic because of redshift of physical momentum.   We
establish a direct connection between the microphysics of decoupling
\emph{in or out of LTE} and the constraints that the particles must
fulfill to be suitable DM candidates \emph{in terms of  the
distribution functions at decoupling}.

\medskip

(\textbf{b:}) We introduce a  primordial coarse grained phase space density
$$
\mathcal{D}  \equiv \frac{n(t)}{\big\langle \vec{P}^2_f \big\rangle^\frac32} \; ,
$$ where $ n(t) $ is the number of particles per unit physical volume
and $\Big\langle\vec{P}^2_f\Big\rangle$ is the average of the
physical momentum with the distribution function of the decoupled
particle. $ \mathcal D $ is a Liouville invariant after decoupling
and only depends on the distribution functions at decoupling. In the
non-relativistic regime $ \mathcal D $ is simply related  to the
phase densities considered in refs.
\cite{dalcanton1,TG,hogan,madsenQ} and   can only {\bf decrease} by
collisionless phase mixing or self-gravity dynamics \cite{theo}.

In the non-relativistic regime we obtain
\be \label{Dint}
\mathcal{D} = \frac1{3^\frac32 \;  m^4} \;
\frac{\rho_{DM}}{\sigma^3_{DM}}
\ee
where $ \sigma_{DM} $ is the
primordial  one-dimensional velocity dispersion and $ \rho_{DM} $
the dark matter density. Combining the result for the primordial
phase space density $ \mathcal D $ determined by the mass and the
distribution function of the decoupled particles, with the recent
compilation of photometric and kinematic data on dSphs satellites in
the Milky-Way \cite{gilmore} yields {\bf lower} bounds on the DM
particle mass $ m $ whereas  {\bf upper} bounds on the DM mass are
obtained using the value of the observed dark matter density today.
Therefore the combined analysis of observational data from (dSphs),
N-body simulations and the present DM density allows us to establish
both \emph{upper and lower} bounds on the mass of the DM candidates.

 We thus provide a link between the microphysics of
decoupling, the observational aspects of dark matter halos and the
DM mass value.

\medskip

(\textbf{c:}) Recent $N$-body simulations \cite{numQ} indicate that
the phase-space density decreases a factor $ \sim 10^2 $ during
gravitational clustering. This result combined with eq.(\ref{Dint})
and the observed values on dSphs satellites \cite{gilmore} yield
 $$
m_{cored}\sim \frac2{g^\frac14} \; \mathrm{keV} \quad , \quad m_{cusp}
\sim\frac8{g^\frac14} \; \mathrm{keV} \; .
$$
for the   masses of \emph{thermal relics} DM candidates, where
`cored' and 'cusp' refer to the type of profile used in the dShps
description and $  1\leq g \leq 4 $ is the number of internal
degrees of freedom of the DM particle. Wimps with masses $ \sim 100
\, \textrm{GeV} $ decoupling in LTE at temperatures $ T_d \sim
10\,\mathrm{MeV} $ lead to primordial phase space densities many
orders of magnitude larger than those observed in (dSphs).  The
results of $N$-body simulations, which yield relaxation by $2-3$
orders of magnitude\cite{numQ} suggest a potential tension for WIMPs
as DM candidates. However, the $N$-body simulations in
ref.\cite{numQ} begin with initial conditions with values of the
phase space density much lower than the primordial one. Hence it
becomes an important question whether the enormous relaxation
required from the primordial values to those of observed in dSphs
can be inferred from numerical studies with suitable (much larger)
initial values of the phase space density.

\medskip

 (\textbf{d:}) We study the possibility that the DM particle is a light Boson that
 undergoes  Bose-Einstein
 Condensation (BEC) prior to decoupling while still ultrarelativistic.
 (This possibility  was addressed in \cite{madsenbec}).
  We analyze in detail the constraints on such BEC DM candidate from
 velocity dispersion and phase space arguments, and contrast the BEC DM
 properties to those of the hot or warm thermal relics.

\medskip

 (\textbf{e:}) Non-equilibrium scenarios that describe various possible
WDM candidates are studied in
 detail. These scenarios describe particle production \cite{boydata} and incomplete
 thermalization \cite{dvd}, resonant \cite{dw} and non-resonant \cite{este}
production of sterile  neutrinos and a model recently proposed
\cite{strigari} to describe cores in dSphs.

 Our analysis of the DM candidates is based on their masses, statistics and  properties
 at decoupling (being it in LTE or not). We combine observations on
 dSphs \cite{gilmore} and $N$-body simulations \cite{numQ}, with
 theoretical analysis using the non-increasing property of the phase
 space density \cite{TG,dalcanton1,hogan,theo}.

The results from the combined  analysis of the primordial phase
space densities, the observational data on dSphs \cite{gilmore} and
the $N$-body simulations in ref.\cite{numQ} are the following:
\begin{itemize}
\item{\textbf{(i):}  conventional
thermal relics, and sterile neutrinos produced resonantly or
non-resonantly  with mass in the range  $ m \sim
\mathrm{few}\,\mathrm{keV} $ that decouple when ultrarelativistic
lead to a primordial phase space density of the same order of
magnitude as in cored dShps and disfavor cusped satellites for which
the data \cite{gilmore} yields a much larger phase space density. }

\item{\textbf{(ii):} CDM from wimps that decouple when non-relativistic with $ m
\gtrsim 100~\mathrm{GeV} $ and kinetic decoupling at $ T_d \sim
10~\mathrm{MeV} $ \cite{dominik} yield phase space densities at
least   eighteen to fifteen orders of magnitude  [see
eqs.(\ref{rss}), (\ref{cusp}) and (\ref{denw})] larger than the
typical  average in dSphs \cite{gilmore}. Results from $N$-body
simulations, albeit with initial conditions with much smaller values
of the phase space density,  yield a dynamical relaxation by a
factor $ 10^2-10^3 $ \cite{numQ}. If these results are confirmed by
simulations with larger initial values there may be a potential
tension between the primordial phase space density for \emph{thermal
relics} in the form of WIMPs with $ m\sim 100\,\mathrm{GeV},  \; T_d
\sim 10 $ MeV  and those observed in dShps.}

\item{\textbf{(iii):}
Light bosonic particles decoupled while ultrarelativistic and which
form a BEC lead to phase space densities consistent with cores and
also consistent with cusps if $ T_c/T_d \gtrsim 10 $. However if
these thermal relics  satisfy the observational bounds,  they must
decouple when $ g_d \; g^{-\frac34} \; (T_d/T_c)^\frac98 > 130 $,
namely {\it above} the electroweak scale.  }
\end{itemize}

\medskip

Section II analyzes the {  generic} dynamics of decoupled particles
for {  any} distribution function, {  with or without} LTE at
decoupling, and for {\bf different} species of particles. In section
III we consider light thermal relics which decoupled in LTE as DM
components: fermions and  bosons, including the possibility of a
Bose-Einstein condensate. Section IV deals with coarse grained phase
space densities which are Liouville invariant and the new bounds
obtained with them by using the observational dSphs data and recent
results from $N$-body simulations, bounds from velocity dispersion,
and the generalized Gunn-Tremaine bound. In Section V we study the
case of particles that decoupled {  out of equilibrium} and the
consequences  on the dark matter constraints. Section VI summarizes
our conclusions.

\section{Preliminaries: dynamics of decoupled particles}

  While the study of kinetics in the early Universe is  available in the
literature \cite{bernstein,kt,scott}, in this section we expand on
the \emph{dynamics of decoupled particles} emphasizing several
aspects relevant to the analysis that follows in the next sections.

Consider a spatially flat FRW cosmology with length element \be
ds^2=dt^2-a^2(t) \; d\vec{x}^2 \ee the non-vanishing Christoffel
symbols are \be \Gamma^0_{ij}=\dot{a} \; a  \; \delta_{ij} \quad ,
\quad \Gamma^i_{0j}=\Gamma^i_{j0} = \frac{\dot{a}}{a} \; \delta^i_j
\,.\ee The (contravariant) four momentum is defined as $ p^\mu =
{dx^\mu}/{d\lambda}$ with $\lambda$ an affine parameter,   so that $
g_{\mu \nu}p^\mu p^\nu = m^2$, where $m$ is the mass of the
particle. This leads to the dispersion relation \be p^0(t) =
\sqrt{m^2+a^2(t) \; {\vec p}^{\,2}(t)} \,. \label{P0} \ee The
geodesic equations are \bea && \frac{dp^0}{d\lambda} = p^0(t)  \;
\dot{p}^0 = - H(t) \;   {a^2(t) \; p^2(t)} \quad  \Rightarrow  \quad
\dot{p}^0 = -H(t) \;  \frac{a^2(t)  \; p^2(t)}{p^0(t)} \label{p0dot}
\cr \cr && \frac{d{\vec p}}{d\lambda} = - 2 \, H(t) \; p^0(t) \;
{\vec p}(t) \quad \Rightarrow \quad \dot{\vec p} = - 2 \, H(t) \;
{\vec p}(t)  \; ,  \label{dotpi} \eea where $ H(t) \equiv
\frac{\dot{a}}{a} $ and we used $ d/d\lambda = p^0 \; d/dt $. The
solution of eq.(\ref{dotpi}) is \be \label{pc} {\vec p} =
\frac{{\vec p}_c}{a^2(t)} \; , \ee where $ p_c $ is the {  time
independent} comoving momentum. The local observables, energy and
momentum as measured by an observer at rest in the expanding
cosmology are given by \be E(t) = g_{\mu \nu} \; \epsilon^\mu_0 \;
p^\nu \quad , \quad P^i_f(t)  =  -g_{\mu \nu} \; \epsilon^\mu_i \;
p^\nu \label{mome} \ee where $ \epsilon^{\mu}_{\alpha} $ form a
local orthonormal tetrad (vierbein)
$$
g_{\mu \nu} \; \epsilon^\mu_\alpha  \; \epsilon^\nu_\beta=\eta_{\alpha
\beta}= \mathrm{diag}(1,-1,-1,-1) \; ,
$$
and the sign in eq.(\ref{mome}) corresponds to a space-like
component. For the FRW metric \be\label{tetrad}
\epsilon^{\mu}_\alpha = \sqrt{\big|g^{\mu \alpha}\big|} \; ,\ee and
we find,
\be
E  =  p^0 \label{E} \quad ,\quad {\vec P}_f(t)   = a(t)
\; {\vec p}(t) = \frac{{\vec p}_c}{a(t)} \; .
\ee
$ \vec{P}_f $ is clearly the physical momentum, redshifting with the expansion.
Combining the above with eq.(\ref{P0}) yields the local dispersion
relation
\be
E(t)= p^0(t)=\sqrt{m^2+{\vec P}^2_f(t)} \label{disp}\; .
\ee
A frozen distribution describing a particle that has been decoupled
from the plasma is constant along geodesics, therefore, taking the
distribution to be a function of the physical momentum $ \vec{P}_f $
and time, it obeys the Liouville equation or collisionless Boltzmann
equation
\be
\frac{d}{d\lambda} f[P_f;t] =0 \quad \Rightarrow \quad
\frac{d f[P_f;t]}{dt}= 0 \; .\label{lio}
\ee
Taking $ P_f $  as an
independent variable this equation leads to the familiar form \be
\label{maslio} \frac{\partial f[P_f;t]}{\partial t}- H(t) \;  P_f \;
\frac{\partial f[P_f;t]}{\partial P_f} = 0 \; . \ee Obviously a
solution of this equation is \be f[P_f;t] \equiv f_d[a(t) \; P_f] =
f_d[p_c] \label{fD} \; , \ee where $ p_c $ is the time independent
comoving momentum. The physical phase space volume element is
invariant, $ d^3X_f \; d^3x P_f = d^3 x_c  \; d^3 p_c $, where $ f,c
$ refer to physical and comoving volumes respectively.

The scale factor is normalized so that \be a(t)= \frac{1+z_d}{1+z(t)} \label{aoft}
\ee and $ P_f(t_d) = p_c $, where $ t_d $ is the cosmic time at
decoupling and $ z $ is the redshift.

\medskip

If a particle
of mass $ m $ has been in LTE but it decoupled from the plasma with
decoupling temperature $ T_d $ its distribution function is
\be
f_d(p_c) = \frac{1}{e^{\frac{\sqrt{m^2+p^2_c}-\mu_d}{T_d}}\pm 1} \;
, \label{LTEdist}
\ee
for fermions $(+)$ or bosons $(-)$
respectively allowing for a chemical potential $ \mu_d $ at decoupling.

In what follows we consider general  distributions  as in
eq.(\ref{fD}) unless specifically stated.

The kinetic energy momentum tensor associated with this frozen
distribution is given by
\be
\label{Tmunu} T^\mu_\nu = g \int
\frac{d^3P_f}{(2\pi)^3} ~\frac{p^\mu  \;  p_{\nu}}{p^0} ~f_d(p_c) \; ,
\ee
where $ g $ is the number of internal degrees of freedom,
typically $ 1 \leq g \leq 4 $. Taking the distribution function to be
isotropic it follows that
\bea
&& T^0_0 = g\int \frac{d^3P_f}{(2\pi)^3} ~ {p^0} ~f_d(p_c)  = \rho \label{rho}\\
&& T^i_j = - \frac{g}{3} \; \delta^i_j     \int
\frac{d^3P_f}{(2\pi)^3} ~ \frac{ a^2(t)  \;  p^2}{p^0} ~f_d(p_c) = -
\frac{g}{3} \;  \delta^i_j \int \frac{d^3P_f}{(2\pi)^3} ~ \frac{
P^2_f}{p^0} ~f_d(p_c) = -\delta^i_j  \; \mathcal{P}  \; , \label{pres}
\eea
where $ \rho $ is the energy density and $ \mathcal{P} $ is the
pressure. In summary,
\be\label{rhopressfin}
\rho = g\int \frac{d^3P_f}{(2\pi)^3}  \;
\sqrt{m^2+P^2_f} ~ f_d[a(t) \; P_f] ~~;~~ \mathcal{P} =
\frac{g}{3}  \;  \int \frac{d^3P_f}{(2\pi)^3}  \;  \frac{
P^2_f}{\sqrt{m^2+P^2_f}}  \; f_d[a(t) \; P_f] \; .
\ee
The pressure can be written in a manner more familiar from kinetic theory as
\be
\mathcal{P} = g \int \frac{d^3P_f}{(2\pi)^3} ~ \frac{ \big|\vec{v}_f\big|^2}{3}~
\sqrt{m^2+P^2_f} \; f_d[a(t) \; P_f]  \; ,
\ee
where $ \vec{v}_f = \vec{P}_f/E $ is the physical (group) velocity of the
particles measured by an observer at rest in the expanding cosmology.

To confirm covariant energy conservation  recall  that $ d^3 P_f =
d^3 p_c/a^3(t); \; P_f = p_c/a(t); \; df_d/dt=0 $, furthermore
from eq.(\ref{p0dot}) it follows that $ \dot{p}^0 = -H(t) \;  P^2_f/p^0 $,
leading to
\be
\dot{\rho} = -3 \, H(t) \; \rho - H(t) \;  g \;
\int \frac{d^3P_f}{(2\pi)^3}  \;  \frac{P^2_f}{p^0}  \; f_d(p_c)  \; ,
\ee
the first term results from the measure and the last term from
$ \dot{p}^0 $; from the expression of the pressure eq.(\ref{pres}) the
covariant conservation equation
\be
\dot{\rho}+3 \, H(t) \;  (\rho+\mathcal{P}) =0 \label{covacons}
\ee
follows.  The number of particles per unit physical volume is
\be
n(t) = g\int \frac{d^3P_f}{(2\pi)^3}
~f_d[a(t) \; P_f] \,,\label{nfin}
\ee
and obeys
\be
\frac{dn(t)}{dt}+3 \, H(t) \; n(t) =0 \,, \label{parnum}
\ee
namely, the number of particles per unit \emph{comoving} volume $ n(t) \;  a^3(t) $ is
conserved.

\medskip

These are   generic  results for the kinetic energy momentum tensor
and the particle density for   any  distribution function that obeys
the collisionless Boltzmann equation (\ref{lio}).

\medskip

The entropy density for an  arbitrary  distribution function for
  particles that decoupled in or out of LTE is
\be s_d(t) = -g\int \frac{d^3P_f}{(2\pi)^3} \Bigg[f_d \ln f_d \pm
(1\mp f_d) \ln (1\mp f_d) \Bigg] \label{entropy} \ee where the upper
and lower signs refer to Fermions and Bosons respectively. Since $
df_d/dt=0 $ it follows that \be \frac{ds_d(t)}{dt} +3  \, H(t)  \;
s_d(t) =0  \; ,\label{endil} \ee therefore the entropy per
\emph{comoving} volume $ s_d(t) \; a^3(t) $  is constant. In
particular the ratio \be Y= \frac{n(t)}{s_d(t)} \ee is a   constant
for \emph{any} distribution function that obeys the collisionless
Liouville equation \cite{kt}.

\medskip

In the case of  LTE, using the  distribution eq.(\ref{LTEdist}) in
the entropy density eq.(\ref{entropy}) yields   the result \be
s_d(t) = \frac{\rho_d + P_d}{T_d  \; a^3(t) }-\frac{\mu_d}{T_d}  \;
n(t) \; , \label{st} \ee for either statistics, where $ \rho_d;P_d $
are evaluated at the decoupling time $ t_d $. The entropy of the gas
of decoupled particles does \emph{not} affect the relationship
between the photon temperature and the temperature of
ultrarelativistic particles that decouple \emph{later} which can be
seen as follows.

Consider several species of particles, one of which decouples at an earlier
time \emph{in or out of equilibrium} with the distribution function
$ f_d $ and entropy given by eq.(\ref{entropy}) while the others remain in
LTE with entropy density $ (2 \, \pi^2/45)  \; g(T)  \; T^3 $,
until some of them decouple later while ultrarelativistic. Here $ T $
is the temperature at time $ t $ and $ g(T) $ is the effective number of ultrarelativistic
degrees of freedom. Entropy conservation leads to the relation,
\be
\left[\frac{2\pi^2}{45}  \; g(T)  \; T^3+ s_d\right] \; a^3(t)
= \mathrm{constant} \; ,
\ee
however, because $ s_d(t)  \; a^3(t) = \mathrm{constant} $, the usual relation
$ g(T) \; T^3 \; a^3(t)=\mathrm{constant} $, relating the temperature $ T $ of a
gas of ultrarelativistic decoupled particles to the photon temperature follows.

\medskip

For light particles that decouple in LTE at temperature $ T_d \gg m $
we can approximate
\be
\frac{\sqrt{m^2+p^2_c}-\mu_d}{T_d} \simeq
\frac{p_c-\mu_d}{T_d} = \frac{P_f -\mu_d(t)}{T_d(t)}
\ee
where
\be
T_d(t) = \frac{T_d}{a(t)}  \quad , \quad \mu_d(t) = \frac{\mu_d}{a(t)}
\label{Tf}
\ee
are the decoupling temperature and chemical potential
red-shifted by the expansion, therefore for particles that decouple
in LTE with $T_d \gg m $ we can approximate
\be \label{dist}
f_d(P_f;t) = \frac1{e^{\frac{P_f-\mu_d(t)}{T_d(t)}}\pm 1} =
\frac{1}{e^{\frac{P_f}{T_d(t)}-\frac{\mu_d}{T_d}}\pm 1}=
\frac{1}{e^{\frac{p_c-\mu_d}{T_d}}\pm 1} \; .
\ee
This distribution function is the same as that of a \emph{massless}
particle in LTE which is also a solution of the Liouville equation,
or collisionless Boltzmann equation.

Since the distribution function is dimensionless, without loss of
generality we can always write   for a particle that
\emph{decoupled} \emph{in or out} of LTE \be f_d(p_c)=
f_d\left(\frac{p_c}{T_d};\frac{m}{T_d};\alpha_i\right) \label{farbi}
\ee where $ \alpha_i $ are  \emph{dimensionless} constants
determined by the microphysics, for example dimensionless couplings
or ratios between $ T_d $ and particle physics scales or in
equilibrium $ \mu_d/T_d $ etc.  To simplify notation in what follows
we will not include explicitly the set of dimensionless constants $
m/T_d, \; \alpha_i $, etc,  in the argument of $ f_d $, but these
are implicit in generic distribution functions. If the particle
decouples when it is  ultrarelativistic, $ m/T_d \rightarrow 0 $.

\medskip

It is convenient to introduce the dimensionless ratios
\be \label{yvar}
y = \frac{p_c}{T_d}= \frac{P_f}{T_d(t)} \quad , \quad
T_d(t)=\frac{T_d}{a(t)}
\ee
and
\be
x_d= \frac{m}{T_d} \quad , \quad
x(t) = \frac{m}{T_d(t)} = a(t) \;  x_d \,.\label{xt}
\ee
For example, for a particle that decouples in equilibrium while
being non-relativistic, $ f_d $ is the Maxwell-Boltzmann distribution
function \cite{kt}
\be \label{MBf}
f_d(p_c) = \frac{2^\frac52 \,\pi^\frac72}{45} \; g_d \; Y_\infty  \;
e^{-\frac{p^2_c}{2 \, m \; T_d}} =
\frac{2^\frac52 \,\pi^\frac72}{45} \; g_d \; Y_\infty \,
e^{-\frac{y^2}{2\,x_d}}  \; ,
\ee
where $ g_d $ is the effective number of ultrarelativistic degrees of freedom at decoupling,
$ Y=n/s $ and $ Y_\infty $ is the solution of the Boltzmann equation, whose
dependence on $ x_d = m/T_d $ and the annihilation cross section is
given in chapter 5.2 in ref. \cite{kt}.

Changing the integration variable in eqs.(\ref{rhopressfin})-(\ref{nfin})
to $ P_f = y   \; T_d(t) $  we find
\bea
&&\rho = g~m~ T^3_d(t) \; I_\rho[x]\quad , \quad
I_\rho[x]=\frac1{2\pi^2}\int^\infty_0 y^2  \; \sqrt{1+
\frac{y^2}{x^2} } \; f_d(y) \; dy \label{rhoy} \cr \cr
&& \mathcal{P} = g~\frac{T^5_d(t) }{3 m} \; I_\mathcal{P}[x] \quad , \quad
I_\mathcal{P}[x]= \frac{1}{2 \, \pi^2}\int^\infty_0 dy \;
\frac{y^4 \; f_d(y)}{\sqrt{1+ \frac{y^2}{x^2}}} =
-x^3  \; \frac{dI_\rho[x]}{dx} \label{presy}\cr \cr
&& n(t)=g~ \frac{T^3_d(t)}{2 \, \pi^2}\int^\infty_0 y^2 \; f_d(y)  \; dy
=  g \; T^3_d(t) \; I_{\rho}[x=\infty]\label{nnoft} \; ,
\eea
 leading to  the equation of state:
\be\label{Wofx}
w[x] = \frac{\mathcal{P}}{\rho} =
\frac{I_\mathcal{P}[x]}{3\,x^2 \; I_\rho[x]} = -\frac13 ~ \frac{d
\ln I_\rho[x]}{d \ln x} \; .
\ee
In the ultrarelativistic and non-relativistic limits, $ x \to 0 $
and $ x \to \infty $, respectively, we find
\bea\label{limi}
&& I_{\rho}[x] \buildrel{x \to 0}\over= \frac1{x} \; \int_0^{\infty}
\frac{y^3 \; dy}{2 \, \pi^2} \; f_d(y) \quad , \quad I_{\mathcal{P}}[x]
\buildrel{x \to 0}\over= x\; \int_0^{\infty}   \frac{y^3 \;
dy}{2 \, \pi^2} \; f_d(y) \; , \cr \cr
&& I_{\rho}[x] \buildrel{x \to
\infty}\over= \int_0^{\infty} \frac{y^2 \; dy}{2\pi^2} \; f_d(y)
\quad , \quad I_{\mathcal{P}}[x] \buildrel{x \to \infty}\over=
\int_0^{\infty} \frac{y^4 \; dy}{2\pi^2} \; f_d(y) \; .
\eea
In the ultrarelativistic limit the energy density and pressure become,
\be
\rho \buildrel{x \to 0}\over= g~  T^4_d(t)    \; \int_0^{\infty}
\frac{y^3 \; dy}{2\pi^2} \; f_d(y) \quad , \quad  \mathcal{P}
\buildrel{x\to 0}\over= \frac{\rho}{3} \quad , \quad  w[x]
\buildrel{x \to 0}\over= \frac13 \quad ,
\ee
describing radiation behaviour. In the non-relativistic limit
\be\label{nonrerhoP}
\rho   \buildrel{x \to \infty}\over= m~g \; T^3_d(t)  \;
\int_0^{\infty} y^2  \;  f_d(y)  \;  \frac{dy}{2\pi^2} \; =m \; n(t)
\quad , \quad \mathcal{P} \buildrel{x \to \infty}\over=
 \frac{g \; T^5_d(t)}{3 \, m} \; \int_0^{\infty} y^4   \; f_d(y) \;
\frac{dy}{2 \, \pi^2} \to 0
\ee
and the equation of state becomes
\be \label{wofxNR}
w[x] \buildrel{x \to \infty}\over=
\frac13 \; \left[\frac{T_d(t)}{m}\right]^2 \frac{\int_0^{\infty} y^4 \;
\; dy  \; f_d(y)}{\int_0^{\infty}  y^2 \; dy  \; f_d(y) } \to 0 \; ,
\ee
corresponding to cold matter behaviour. In the
non-relativistic limit, it is convenient to write
\be
\rho = m \, n_\gamma(t) \; g  \; \Bigg[\frac{T_d(t)}{T_{\gamma}(t)}\Bigg]^3
\frac{\int^\infty_0 y^2 \; f_{d,a}(y) \; dy}{4 \,\zeta(3)} = m \,
n_\gamma(t)\,\frac{g\,\int^\infty_0 y^2  \; f_{d,a}(y) \; dy}{2\,g_d\,
\, \zeta(3)} \; , \label{rhorel}
\ee
where $ \zeta(3) = 1.2020569\ldots, \;  g_d $ is the number of
ultrarelativistic degrees of freedom at decoupling, and $ n_\gamma(t) $ is
the photon number.

The average squared velocity of the particle is given in the non-relativistic limit by
\be
\Big\langle \vec{V}^2 \Big\rangle = \Big\langle
\frac{\vec{P}^2_f}{m^2} \Big\rangle = \frac{\int
\frac{d^3P_f}{(2\pi)^3}  \;  \displaystyle{\frac{\vec{P}^2_f}{m^2}} \; f_d[a(t)P_f]}{\int
\frac{d^3P_f}{(2\pi)^3} \;
f_d[a(t)P_f]} = \left[\frac{T_d(t)}{m}\right]^2  \; \frac{\int_0^\infty y^4 f_d(y) dy}{\int_0^\infty y^2
 f_d(y)dy} \; . \label{vel2}
\ee
Therefore, the equation of state in thermal equilibrium is given by
\be
\mathcal{P}  = \frac13 \; \Big\langle \vec{V}^2 \Big\rangle \,
\rho \equiv \sigma^2  \; \rho \quad , \quad
 \sigma= \sqrt{\frac13  \; \Big\langle \vec{V}^2 \Big\rangle} \; ,
 \label{isoeqnofstate}
\ee
where $ \sigma $ is the \emph{one dimensional} velocity dispersion  given  at redshift $ z $ by
\be \label{sigdmz}
\sigma(z) =  \frac{T_d(t)}{m \; }\, \Bigg[ \frac{\int_0^\infty y^4  \; f_d(y) \;
dy}{3\,\int_0^\infty y^2 \;
 f_d(y) \; dy}\Bigg]^{\frac12}=0.05124 \; \frac{1+z}{g^{\frac13}_d}\ \; \Big(\frac{\mathrm{keV}}{m}\Big)  \;
  \Bigg[ \frac{\int_0^\infty y^4  \; f_d(y)  \; dy}{ \int_0^\infty y^2 \;
 f_d(y) \; dy}\Bigg]^{\frac12} \; \Big(\frac{\mathrm{km}}{\mathrm{s}}  \Big) \; .
\ee
and we used that
\be \label{tdt}
T_d(t) = T_d \; (1+z) = \left(\frac2{g_d}\right)^\frac13 \; T_{\gamma} \; (1+z) \; ,
\ee
$ T_{\gamma} = 0.2348 \times 10^{-3} \,\mathrm{eV}$  is the photon
temperature today \cite{pdg}.

\medskip

The results above, eqs.(\ref{rhoy})-(\ref{sigdmz}) are   general
for any distribution of decoupled particles {\bf whether or not} the
particles decoupled in equilibrium.

Using the relation
(\ref{rhorel}) for a given species $ (a) $ of particles with $ g_a $
degrees of freedom, their relic abundance \emph{today} is given by
\be \label{relic}
\Omega_a \; h^2 = \frac{m_a}{25.67~\mathrm{eV}} \;  \frac{g_a \;
\int^\infty_0 y^2  \; f_{d,a}(y) \; dy}{2 \;  g_{d,a} \; \zeta(3)} \; .
\ee
where we used that today $ h^2 \; n_\gamma/\rho_c = 1/25.67 $eV \cite{pdg}.

If this decoupled species contributes a fraction $ \nu_a $ to dark matter,
with $ \Omega_a = \nu_a \; \Omega_{DM} $  and using that $ \Omega_{DM} \; h^2 =
0.105 $ \cite{pdg} for non-baryonic dark matter, then:
\be
\nu_a = \frac{m_a}{2.695~\mathrm{eV}} \; \frac{g_a\int^\infty_0 y^2 \;
f_{d,a}(y) \; dy}{2  \; g_{d,a} \; \zeta(3)} \; . \label{nua}
\ee
Since $ 0\leq \nu_a \leq 1 $ we find the  constraint
\be
m_a \leq 2.695 \; \mathrm{eV} \; \frac{2  \; g_{d,a} \; \zeta(3)}{g_{a} \;
\int^\infty_0 y^2  \; f_{d,a}(y) \; dy} \; , \label{bound}
\ee
where in general $ f_d $ depends on the mass of
the particle as in eq.(\ref{farbi}). For a particle that decouples
while non-relativistic with the distribution function eq.(\ref{MBf})
this is recognized as the generalization of the Lee-Weinberg {\bf lower
bound} \cite{LW,kt}, whereas if the particle decouples in or out of
LTE when it is ultrarelativistic, in which case $ f_{d,a}(y) $ \emph{does
not} depend on the mass, eq.(\ref{bound}) provides and {\bf upper bound}
which is a generalization of the Cowsik-McClelland \cite{cow,kt}
bound.

The constraint eq.(\ref{bound}) suggests \emph{two} ways to allow for
more massive particles: by increasing $ g_d $, namely the particle
decouples earlier, at higher temperatures when the effective number
of   ultrarelativistic species is larger, and/or decoupling out of
LTE with a distribution function that favors \emph{smaller} momenta,
thereby making the denominator in eq.(\ref{bound}) smaller, the smaller
number of particles allows a larger mass to saturate the DM abundance.

\medskip

For the particle to be a suitable dark matter candidate, the free
streaming length must be much smaller than the Hubble radius.
Although we postpone to a companion article \cite{free} a more
detailed study of the free streaming lengths in terms of the
\emph{generalized distribution functions}, here we adopt the simple
requirement that the velocity dispersion   be small, namely the
particle must be non-relativistic
\be
\langle \vec{V}^2 \rangle =
\big\langle \frac{\vec{P}^2_f}{m^2} \big\rangle \ll 1 \,.\label{NR}
\ee From eq.(\ref{vel2}) this constraint yields \be \frac{m}{T_d(t)}
\gg \sqrt{\frac{\int_0^\infty y^4 \;  f_d(y) dy}{\int_0^\infty y^2
\; f_d(y)dy}} \; , \label{xz}
\ee
where $ T_d(t) $ is given by
eq.(\ref{yvar}). From eqs.(\ref{aoft}), (\ref{xt}), (\ref{tdt}) and
(\ref{xz}) we obtain the following condition for the particle to be
non-relativistic at redshift $ z $
\be\label{mNR}
m \gg 2.958  \; \frac{1+z}{g^{\frac13}_d} \times 10^{-4} \;
\sqrt{\frac{\int_0^\infty y^4 \;  f_d(y) \; dy}{\int_0^\infty y^2 \;
f_d(y) \; dy}}  \; \mathrm{eV}  \; .
\ee
Taking the relevant value
of the redshift  for  large scale structure to be the redshift at
which reionization occurs $ z_s \sim 10 $ \cite{wmap3}, we find the
following \emph{generalized} constraint on the mass of the particle
of species $(a)$ which is a dark matter component
\be
\label{marange} \frac{2.958}{g^{\frac13}_d} \times 10^{-4}
\sqrt{\frac{\int_0^\infty y^4  \;  f_{d }(y) \; dy}{\int_0^\infty
y^2 \;  f_{d }(y) \; dy}} \; \mathrm{eV} \ll m  \leq 2.695 \;
\frac{2 \, g_d \,\zeta(3)}{g \int^\infty_0 y^2 \; f_{d }(y) \; dy}
\; \mathrm{eV} \; .
\ee
The left side of the inequality corresponds
to the requirement that the particle be non-relativistic at
reionization (taking $ z_s \sim 10 $), namely a small velocity
dispersion $ \langle \vec{V}^2/c^2 \rangle \ll 1 $, corresponding to
a free streaming length $ \lambda_{fs} \sim  \sqrt{\langle
\vec{V}^2/c^2 \rangle}~ d_H $  much smaller  than the Hubble radius
($d_H$), while the right hand side is the constraint from the
requirement that the decoupled particle is a dark matter
\emph{component}, namely eq. (\ref{bound}) is fulfilled.

\section{Light Thermal relics as Dark Matter Components.}
In this section we consider  particles that decouple in LTE.

\subsection{ Fermi-Dirac and non-condensed Bose-Einstein gases of light particles
as DM components.}

\begin{figure}[h]
\begin{center}
\includegraphics[height=3in,width=3in,keepaspectratio=true]{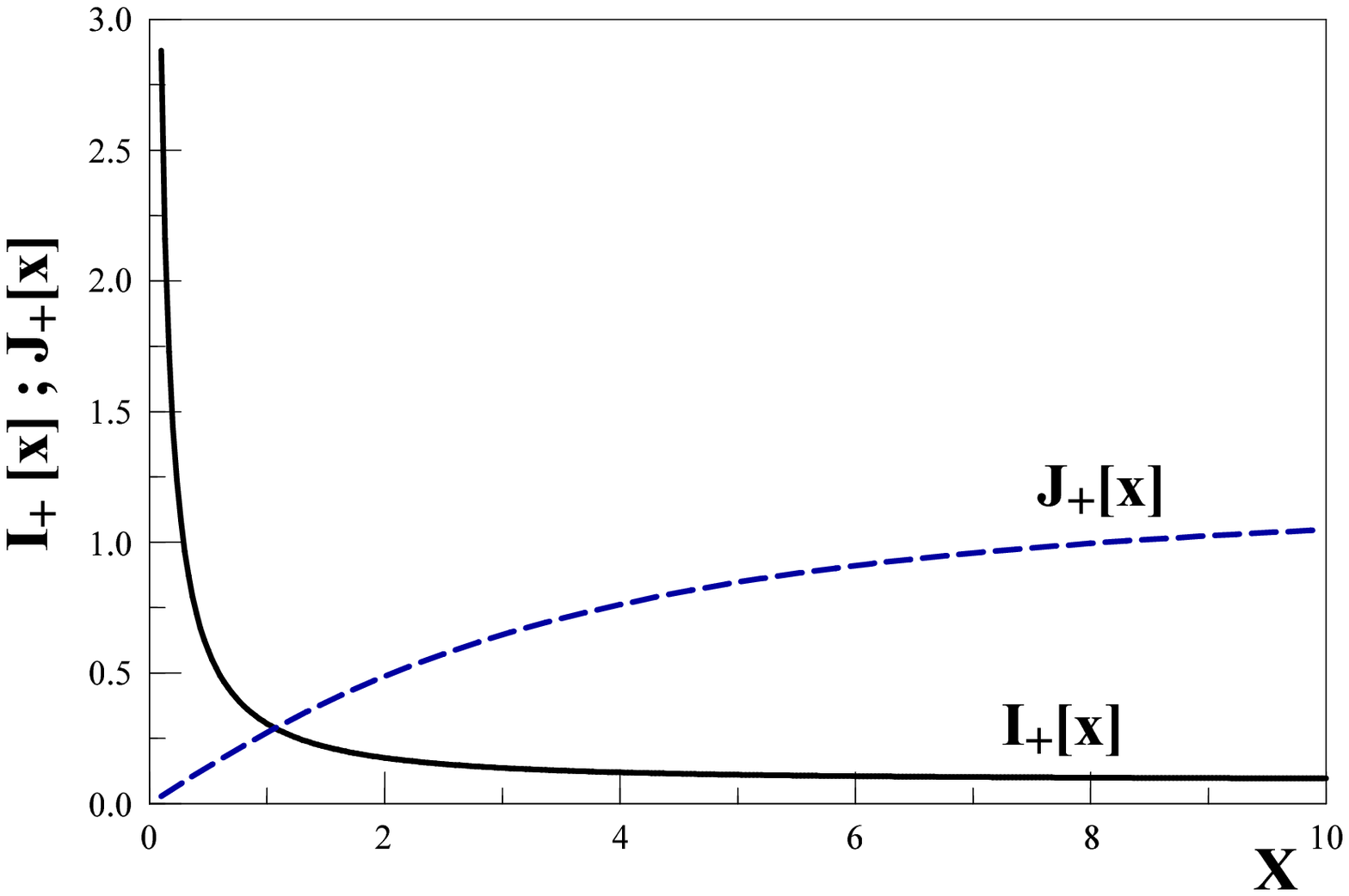}
\includegraphics[height=3in,width=3in,keepaspectratio=true]{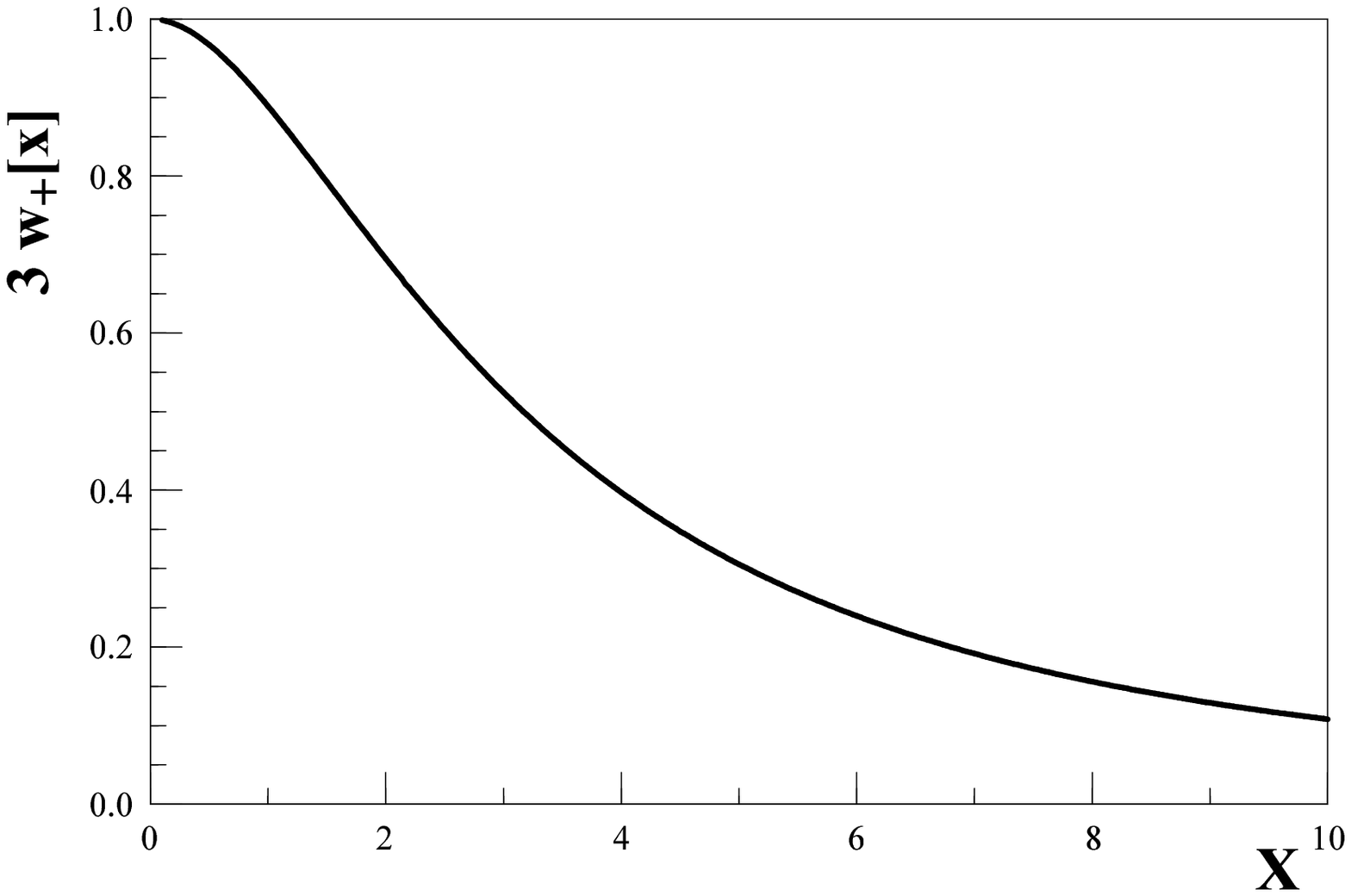}
\caption{Fermions without chemical potential.
Left panel: $ I_+(x) $ and $ J_+(x) $ vs $ x $. Right panel: $ 3 \; w[x] $
vs. $ x. \; I_+ = I_\rho, \; J_+=I_\mathcal{P} $.}
\label{fig:fermions}
\end{center}
\end{figure}

\begin{figure}[h]
\begin{center}
\includegraphics[height=3in,width=3in,keepaspectratio=true]{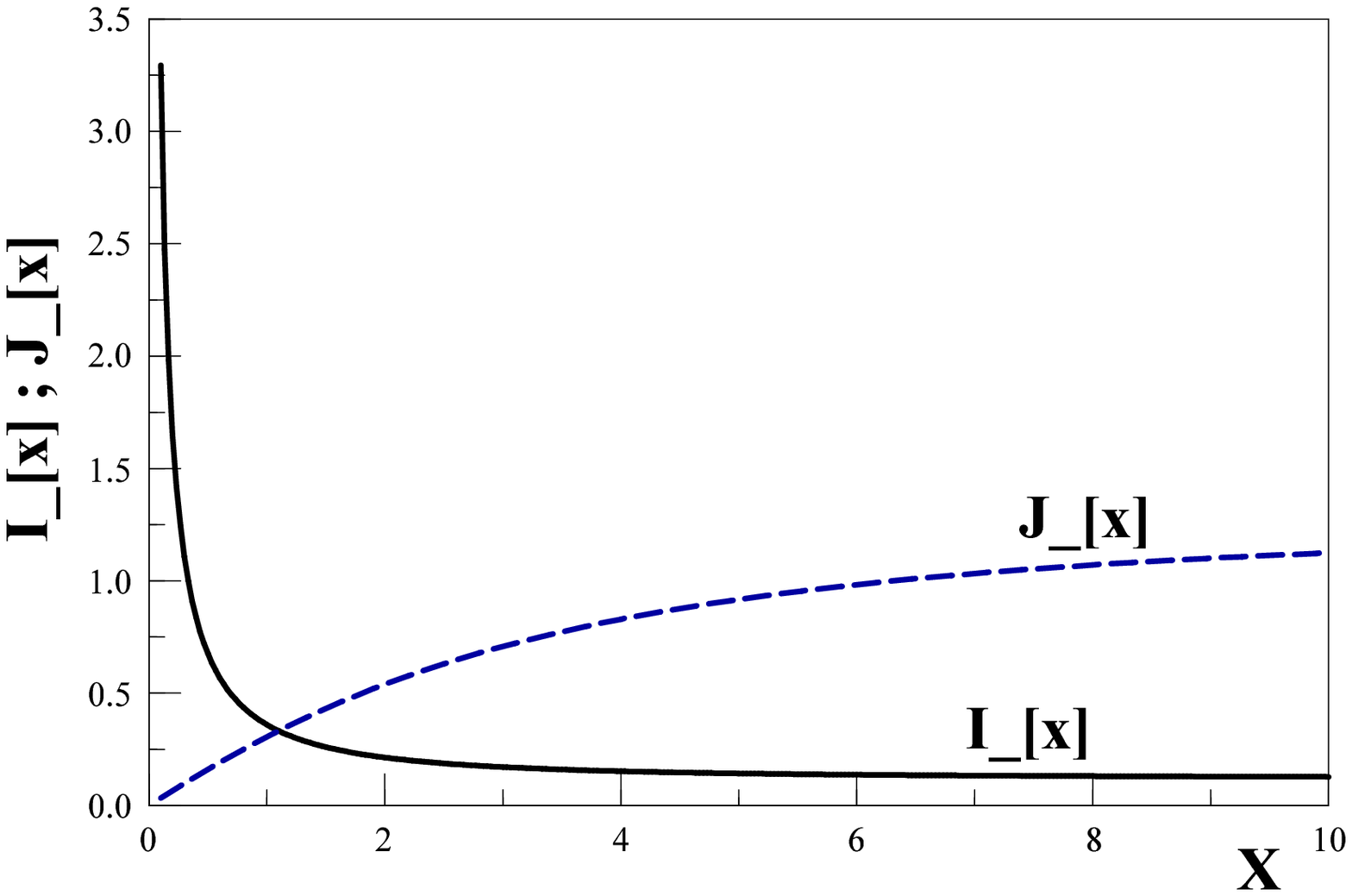}
\includegraphics[height=3in,width=3in,keepaspectratio=true]{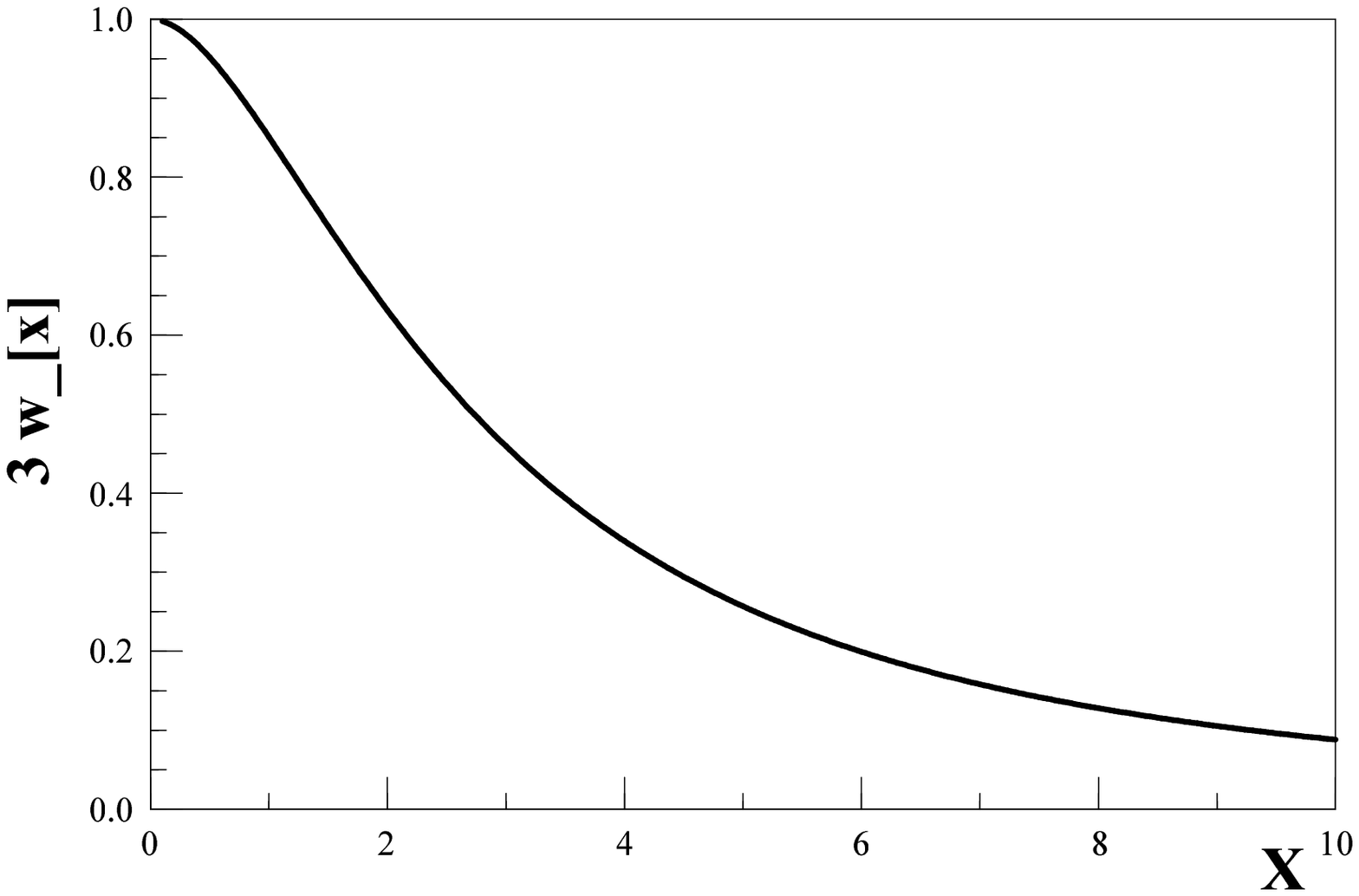}
\caption{Bosons without chemical potential.
Left panel: $ I_-(x) $ and $ J_-(x) $ vs $ x $. Right panel: $ 3 \; w[x] $
vs. $ x . \; I_- = I_\rho , \; J_-=I_\mathcal{P} $. } \label{fig:bosons}
\end{center}
\end{figure}

The functions $ I_\rho(x), \; I_\mathcal{P}(x) $ in the density and pressure
denoted by $ I_\pm(x), \; J_\pm(x) $ respectively for Fermions ($+$) and Bosons $(-)$
and the equation of state $ w[x] $ eq.(\ref{Wofx}) for each case are depicted in
figs. \ref{fig:fermions}-\ref{fig:bosons} for vanishing chemical
potential in both cases. We have also numerically studied these
functions for values of the chemical potential in the range $ 0 \leq
|\mu_d|/T_d \leq 0.5 $ but the difference with the case of vanishing
chemical potentials is less than $ \sim 5\% $ even for the largest
value studied  $ |\mu_d|/T_d =1.0 $ which is about the maximum
consistent with constraints on lepton asymmetries allowed by BBN and
CMB \cite{kneller}.

These figures make clear that the onset of the non-relativistic
behavior occurs for $ x_{nr} \sim 5 $ in both cases. It is useful to
compare this result, with the generalized constraint eq.(\ref{xz}) for
the case of thermal relics. Replacing the LTE distribution functions
(Fermi-Dirac or Bose-Einstein, without chemical potentials) in
eq.(\ref{xz})  we obtain
\be\label{trx}
x_{nr} >   3.597 ~~\mathrm{for~Fermions} \quad , \quad
x_{nr}  >   3.217 ~~\mathrm{for~Bosons} \; .
\ee
The detailed analysis
of the corresponding functions yields the more precise estimate
$ x_{nr} \gtrsim 5 $ in both cases for the transition to the
non-relativistic regime.

Therefore, the decoupled particle of mass $m$ becomes
non-relativistic at a time $ t^* $ when $ m \gtrsim 5\, T_d(t^*) $. At
the time of Big Bang Nucleosynthesis (BBN) when \cite{kt} $ T_{BBN}
\sim 0.1 $MeV and $ g_{BBN} \sim 10 $, the decoupled particle is
non-relativistic if
\be
m \gtrsim g^{-\frac{1}{3}}_d \; \mathrm{MeV} \; ,
\label{mBBN}
\ee
in which case it does not contribute to the
effective number of ultrarelativistic degrees of freedom during BBN and
would not affect  the primordial abundances of light elements. If
the particle remains ultrarelativistic during BBN the total energy
density in radiation is \cite{kt}
\be\label{rhotot}
\rho_{tot}(t) = \frac{\pi^2}{30} \; g_{*}(t) \; T^4_\gamma(t) \left [
1+   \frac{c \; g}{g_{*}(t)}\,\Bigg(\frac{g_{*}(t)}{g_d}
\Bigg)^{\frac43} \right] \; ,
\ee
where $ T_\gamma $ is the (LTE) temperature of the fluid, $ c=1 (7/8) $  for Bosons
(Fermions), $ g_{*}(t) $ is the effective number of ultrarelativistic
degrees of freedom at time $ t $ from particles that \emph{remain} in
LTE at this time, and $ g_d $ is the effective number of degrees of
freedom at decoupling. The second term in eq.(\ref{rhotot}) is an
extra contribution to the effective number of ultrarelativistic degrees
of freedom.

At the time of BBN, $ g_{*}(t_{BBN}) \sim 10 $ \cite{kt}
and early decoupling of the light particle, $ g_d \gg
g_{*}(t_{BBN}) $, leads to a negligible contribution to the effective
number of ultrarelativistic degrees of freedom well within the
current bounds \cite{verde}. Therefore, provided that the decoupled
particle is {\it stable}, either for light particles that remain
relativistic during (BBN) but that decouple very early on when $ g_d
\gg 10 $  or when the particle's mass $ m > 1 $MeV, there is
{\it no} influence on the primordial abundance of light elements and BBN
does {\it not} provide any tight constraints on the particle's mass.

\subsection{A Bose condensed light particle as a Dark Matter component}

Consider the case of a light bosonic particle, for example an
axion-like-particle. Typical interactions involve two types of
processes, inelastic reactions are number-changing processes and
contribute to chemical equilibration, while elastic ones distribute
energy and momenta of the intervening particles, these do not change
the particle number but lead to kinetic equilibration. Consider the
case in which chemical freeze out occurs \emph{before} kinetic
freeze-out, such is the case for a real scalar field with quartic
self-interactions. In this theory, number-conserving processes such
as $ 2 \leftrightarrows 2 $ establish kinetic (thermal) equilibrium,
but conserve particle number, a cross section for such process is
$\propto \lambda^2$ where $\lambda$ is the quartic coupling. The
lowest order number-changing processes that contribute to chemical
equilibrium are $ 4 \leftrightarrows 2 $, with cross sections $ \propto
\lambda^4 $. Hence this is an example of a theory in which chemical
freeze out occurs well before kinetic freeze out for small coupling.

Another  relevant example is the case of WIMPs     studied in ref.
\cite{dominik} where it was found that $ T_{cd}\sim 10\,\mathrm{GeV}
$, while $ T_{kd}\sim 10\,\mathrm{MeV} $ where $ T_{cd},T_{kd} $ are
the chemical and kinetic (thermal) decoupling temperatures
respectively.  Although this study focused on a fermionic particle,
it is certainly possible that a similar situation, namely chemical
freeze-out much earlier than kinetic freeze out, may arise for
bosonic DM candidates.

Under this circumstance, the number of particles is conserved if the
particle is stable, but the temperature continues to redshift by the
cosmological expansion, therefore the gas of Bosonic particles cools
at constant comoving particle number. This situation must eventually
lead to Bose Einstein condensation (BEC) since the thermal
distribution function can no longer accomodate the particles with
non-vanishing momentum within a thermal distribution. Once thermal
freeze out occurs, the frozen distribution \emph{must} feature a
homogeneous condensate and the number of particles for zero momentum
becomes macroscopically large. Although some aspects of Bose
Einstein condensates were studied in ref. \cite{madsenQ,madsenbec},
we study new aspects  such as the impact of the BEC upon the bound
for the mass and \emph{the velocity dispersion} of DM candidates.

 The bosonic distribution function for a fixed number of particles
includes a chemical potential and is given by eq.(\ref{LTEdist})
where $ \mu_d \leq m$ for the distribution function to be manifestly
positive for all $ p $. Separating explicitly the contribution from
the $ \vp =0 $ mode the number of particles \emph{per comoving} volume
$ V_c $ is
\be \label{cond}
n= {\frac{1}{V_c} \frac{1}{e^{\frac{m-\mu_d}{T_d}}-1}}+
\frac{1}{V_c} \sum_{\vp_c}
\frac{1}{e^{\frac{\sqrt{m^2+p^2_c}-\mu_d}{T_d}}- 1}\equiv n_0 + \int
\frac{d^3p_c}{(2\pi)^3}
\frac{1}{e^{\frac{\sqrt{m^2+p^2_c}-\mu_d}{T_d}}- 1}\ee where \be n_0
= {\frac{1}{V_c} \frac{1}{e^{\frac{m-\mu_d}{T_d}}-1}}
\ee
is the comoving condensate density. In the infinite
volume limit the condensate term vanishes unless $ \mu_d \rightarrow
m $. For $ m/T_d \ll1 $ we find
\be
n= n_0 + \frac{T^3_d \;
\zeta(3)}{\pi^2}\,Z\left[e^{\frac{\mu_d}{T_d}}\right] \label{rela}
\ee
where
\be
Z[e^{\frac{\mu_d}{T_d}}] = \frac{1}{\zeta(3)}\sum_l^\infty
\frac{e^{\frac{l \mu_d}{T_d}}}{l^3}\,.
\ee
The maximum value that $ \mu_d $ can achieve is $ m $, therefore,   neglecting $ m/T_d $ we
replace $ Z[e^{\frac{\mu_d}{T_d}}] $ by   $ Z[1]=1 $. If the comoving particle density
\be
n > \frac{T^3_d \, \zeta(3)}{\pi^2}
\label{BECcond}
\ee
then, there must be a zero momentum condensate
with $ n_0 \neq 0 $ and  $ \mu_d =m $ in the infinite (comoving)
volume limit. In this limit we find,
\be
1-\frac{n_0}{n} = \Bigg\{\begin{array}{c}
          \Big(\frac{T_d}{T_{c}} \Big)^3 \quad {\rm for} \quad T_d < T_{c} \\
          0 \quad {\rm for} \quad T_d>T_{c}
        \end{array} \label{fract}\ee
where the critical temperature is given by
        \be
T_{c} = \Big[ \frac{\pi^2 \;  n}{\zeta(3)}\Big]^{\frac{1}{3}}\,. \label{TC}
\ee
The solution of the equation (\ref{cond}) that determines the condensate
fraction shows that for $ T_d<T_{c} $
\be
\mu_d = m \,.\label{condi}
\ee
In the infinite volume limit the distribution function for
particles that decouple while ultrarelativistic $ m/T_d\ll 1 $,  for $ T_d<T_{c} $
becomes
\be
f_d(p_c) = n_0 \, \delta^{(3)}(\vp_c)+
        \frac{1}{e^{\frac{p_c}{T_d}}-1} \,.\label{distri}
\ee  From eq.(\ref{nfin}) the total number of particles for $
m/T_d\ll 1, \; T_c> T_d $ is given by \be n(t)= n_0(t) +
\frac{\zeta(3)}{\pi^2} \; T^3_d(t) \; , \label{nbos}\ee where \be
n_0(t) = \frac{n_0}{a^3(t)}\,.
        \label{condex}
\ee
For $ T_d < T_{c} $ eq.(\ref{fract}) implies that
\be \label{n0f}
n_0(t) =\frac{\zeta(3)}{\pi^2}\left[\left(\frac{T_c}{T_d}\right)^3-1\right]
\; T^3_d(t) \; ,
\ee
hence for $ T_d<T_c $ the total density is given by
\be \label{totn}
n(t) = \frac{\zeta(3)}{\pi^2} \left(\frac{T_c}{T_d}\right)^3 \; T^3_d(t) \; .
\ee
The enhancement factor $ (T_c/T_d)^3 $ over the thermal result reflects the population
of particles in the condensed, zero momentum state. The energy density and pressure are given by
\bea
\rho(t) & = &  g~m
\left\{ n_0(t) + T^3_d(t) \; I^{nc}_\rho[x(t)] \right\}
\label{rhobos}\\
\mathcal{P}(t) & = &   g~\frac{T^5_d(t) }{3 \, m} \;
I^{nc}_\mathcal{P}[x(t)]\label{Pbos}
\eea
where
\bea
I^{nc}_\rho[x(t)] & = &  \frac{1}{2 \, \pi^2} \int_0^\infty
\sqrt{1+\frac{y^2}{x^2(t)}} \; \frac{y^2}{e^y -1}
~dy  \label{Inc}\\
I^{nc}_\mathcal{P}[x(t)] & = &
\frac{1}{2 \, \pi^2}\int^\infty_0 \frac{y^2}{\sqrt{1+ \frac{y^2}{x^2(t)}}}
\; \frac{y^2}{e^y -1} \; dy  \quad , \quad
x(t) = \frac{m}{T_d(t)}  \; ,
\eea
are the contributions from the particles outside the condensate ($  p\neq  0 $).

Two important aspects emerge from these expressions: i) \emph{the
condensate always contributes as a  non-relativistic  component},
ii) the condensate \emph{does not} contribute to the pressure.

\begin{figure}[h]
\begin{center}
\includegraphics[height=3in,width=3in,keepaspectratio=true]{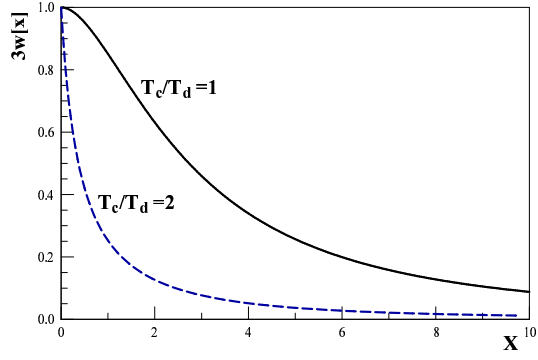}
\caption{$ 3 \; w[x] $ vs $ x $ for the Bose condensed case for
$ T_c/T_d=1-2 $.} \label{fig:3wbec}
\end{center}
\end{figure}

Replacing eq.(\ref{n0f}) into (\ref{rhobos}) and using
$$
\int^\infty_0 \frac{y^2 \; dy}{e^y-1} = 2 \, \zeta(3) \; ,
$$
the energy density and equation of state for $ T_d< T_c $ can be written compactly as
\be
\rho(t) = g~m~ T^3_d(t)~ \mathcal{I}[x(t)] \label{rhocom}
\ee
where
\be \label{Ioft}
\mathcal{I}[x(t)]=  \frac{1}{2\pi^2} \int_0^\infty
\Bigg\{\left[\left(\frac{T_c}{T_d}\right)^3-1\right]+
\sqrt{1+\frac{y^2}{x^2(t)}} \Bigg\}\,\frac{y^2}{e^y -1} \; dy \; .
\ee
The equation of state
\be \label{wbos}
w[x] = \frac{I^{nc}_\mathcal{P}[x] }{3 \, x^2 \; \mathcal{I}[x]}
\ee
is displayed in fig. \ref{fig:3wbec}, from which
it is clear that for $ T_c/T_d >1 $ the non-relativistic limit sets in
{\it much earlier} than for the non-Bose condensed case. This is a
consequence of the zero momentum particles in the BEC which
contribute as \emph{pressureless} cold matter, even when the light
bosonic particle decouples while ultrarelativistic.

For  $ T_d < T_{c} $ when the particle becomes non-relativistic, namely
$ x\rightarrow \infty $,  the energy density becomes
\be
\rho(t) = g~ m~ n(t)
\ee
where $ n(t) $ is the total number of particles per physical volume, including
the condensate and non-condensate components, from eq.(\ref{totn}) it follows that
\be \label{rhobel}
\rho(t) = g~m~ \frac{\zeta(3)}{\pi^2}\left(\frac{T_c}{T_d}\right)^3 \; T^3_d(t) \; ,
\ee
from which for $ T_c\geq T_d $ it follows analogously to eq.(\ref{relic}) that
\be\label{Omeb}
\Omega_{BE} \; h^2 =  \frac{m}{25.67~\mathrm{eV}} \;
\frac{g}{g_d} \left(\frac{T_c}{T_d}\right)^3 \; .
\ee
The dark matter fraction that these particles can contribute is given by
\be
\nu_{BE} = \frac{m }{2.695~\mathrm{eV}} \; \frac{g}{g_d}
\left(\frac{T_c}{T_d}\right)^3 \; ,\label{fracbos}
\ee
resulting in the \emph{upper} bound
\be
m  \leq  2.695   \;  \frac{g_d}{g} \;
\left(\frac{T_d}{T_c}\right)^3 ~  \mathrm{eV} \; .\label{boundBEC}
\ee
In the Bose condensed case $ T_d/T_c <1 $ the bosonic particle is light unless it decouples very
early on at high temperature with a large $g_d$.
The presence of a BEC \emph{tightens} the constraint on the
mass of the light bosonic particle via the extra factor
$ \left( T_d/T_c\right)^3 $ in (\ref{boundBEC}).

Of importance for clustering, is the velocity dispersion when the particle
becomes non-relativistic, it is given by
\be
\Big\langle \vec{V}^2   \Big\rangle =  \Big\langle \frac{\vec{P}^2_f}{m^2} \Big\rangle =  12 \;
\frac{\zeta(5)}{\zeta(3)}\, \Bigg[\frac{T_d(t)}{m} \Bigg]^2 \, \Big(\frac{T_d}{T_c}
\Big)^3 \label{veldispbec}
\ee
where $ \zeta(5)= 1.0369278\ldots $.

The presence of the BEC,  accounted for by the factor $ \Big(
T_d/T_c \Big)^3 <1 $ in eq.(\ref{veldispbec}),  \emph{diminishes} the
velocity dispersion. This is a consequence of the fact that the
particles in the condensate all have vanishing momentum, and only
the non-condensate particles contribute to the velocity dispersion
but the fraction of particles outside of the condensate is precisely
the factor $ \Big( T_d/T_c \Big)^3 $. Therefore the presence of a
BEC leads to a \emph{decrease}  in the velocity dispersion and
consequently \emph{even for light particles to a decrease in the
free streaming length}.

These results imply that although the Bosonic particle is bound to
be very light   by the bound (\ref{boundBEC}) (unless they decoupled
very early ), if $ T_c \gg T_d $ it is \emph{not} a HDM component
but can effectively act as either a WDM or CDM because of a \emph{small
velocity dispersion}. Whether $ T_c \gg T_d $ or not has to be
studied within the microscopic particle physics model that describes
this DM component.

\section{Coarse grained phase space densities and new DM Bounds}

In their seminal article Tremaine and Gunn \cite{TG} argued that the
coarse grained phase space density is always smaller than  or equal
to the maximum of the (fine grained) microscopic phase space
density, which is the distribution function. Such argument relies on
the theorem \cite{theo} that states that collisionless phase mixing
or violenty relaxation by gravitational dynamics can only diminish
the coarse grained phase space density. A similar argument was
presented by Dalcanton and Hogan \cite{dalcanton1,hogan}, and
confirmed by recent numerical studies \cite{numQ}.

As noticed in
ref. \cite{madsenQ}, the case of the Bose-Einstein distribution,
requires a careful treatment because for massless particles the
Bose-Einstein distribution diverges at small momentum. This
divergence is present if there is a BEC even when the mass of the
Bosonic particle is included. This is so since $ \mu_d=m $ is
required to form a BEC and the distribution functions diverge at
zero momentum, even the part of the distribution function that
describes the particles outside the condensate diverges at $ P = 0
$. Madsen recognized this caveat in the Bosonic case and in ref.
\cite{madsenQ} introduced an alternative \emph{statistical
interpretation} of the phase space density, similar to that
introduced in \cite{dalcanton1,hogan} but with the upper limit in
the momentum integrals replaced by a (physical) momentum cutoff as
suggested by the phase mixing theorem \cite{theo}. However, it is
straightforward to show that the resulting coarse grained phase
space density is \emph{not} a Liouville invariant. Instead, we
define the coarse grained (dimensionless) primordial phase space
density
\be
\mathcal{D}  \equiv \frac{n(t)}{\big\langle \vec{P}^2_f
\big\rangle^\frac32} \; , \label{D}
\ee
which is Liouville invariant
and where $ \big\langle \vec{P}^2_f \big\rangle $ is defined   in
eq.(\ref{vel2}). Since the distribution function is frozen and is a
solution of the collisionless Boltzmann (Liouville) equation
(\ref{maslio})  it is clear that $ \mathcal{D} $ is a
\emph{constant}, namely a Liouville invariant in absence of
self-gravity,  . Including explicitly a possible BEC, $ \mathcal{D}
$ is given by \be \displaystyle \mathcal{D} = \frac{ g}{2\pi^2}
\frac{\Bigg[\frac{2\pi^2\,n_0}{T^3_d}+\int_0^\infty y^2 f_d(y) dy
\Bigg]^{\frac{5}{2}}}{\Bigg[\int_0^\infty y^4 f_d(y)dy
\Bigg]^{\frac{3}{2}}}\; , \label{Dex} \ee where $ f_d(y) $ is the
distribution function for the non-condensed particles in the Bosonic
case and $ n_0 $ is the \emph{comoving} density of the Bose-Einstein
condensate \be \frac{2 \, \pi^2 \; n_0}{T^3_d} = \left\{
\begin{array}{l}
          2\,\zeta(3)\left[\left(\frac{T_c}{T_d}\right)^3-1\right]~~
\mathrm{for ~ the ~ BEC ~ with} ~ T_d<T_c \\
        0 ~~ ~~ ~~~\mathrm{for~ the ~Fermionic ~or ~non-Bose~ condensed~
        case \; .}
       \end{array} \right.
\ee
When the particle becomes non-relativistic $ \rho(t) = m  \; n(t) $
and $ \big\langle \vec{V}^2 \big\rangle = \big\langle
\frac{\vec{P}^2_f}{m^2} \big\rangle $, therefore,
\be
\mathcal{D} = \frac{\rho}{m^4\,\big\langle \vec{V}^2 \big\rangle^{\frac{3}{2}} } =
\frac{Q_{DH}}{m^4}\label{DNR}
\ee
where $ Q_{DH} $ is the phase-space
density introduced by Dalcanton and Hogan \cite{dalcanton1,hogan}
\be
Q_{DH}= \frac{\rho}{\big\langle \vec{V}^2 \big\rangle^{\frac32}}
\label{QDH}
\ee
and the one-dimensional velocity dispersion $ \sigma $ is defined by eq.(\ref{isoeqnofstate}).

In the non-relativistic regime $ \mathcal{D} $
is related to the coarse grained phase space density $ Q_{TG} $ introduced by
Tremaine and Gunn \cite{TG}
\be \label{QTG}
Q_{TG}= \frac{\rho}{m^4 \; (2 \, \pi \; \sigma^2)^\frac32} =  \left( \frac3{2 \, \pi}
\right)^\frac32 \; \mathcal{D} \; .
\ee
The observationally accessible quantity is the phase space density
$ \rho/\sigma^3 $, therefore, using  $ \rho = m \;  n $  for a decoupled
particle that is non-relativistic today and eq.(\ref{isoeqnofstate}),
we define the primordial phase space density
\be
\frac{\rho_{DM}}{\sigma^3_{DM}} = 3^\frac32 \;  m^4  \; \mathcal{D}
\equiv 6.611 \times 10^8  \;  \mathcal{D} \;
\Big[\frac{m}{\mathrm{keV}}\Big]^4  \;
\frac{M_\odot/\mathrm{kpc}^3}{\big(\mathrm{km}/\mathrm{s}\big)^3}
 \; \,. \label{PSDM}
\ee
where we used that $ \mathrm{keV}^4 \; \big(\mathrm{km}/\mathrm{s}\big)^3 =
1.2723 \; 10^8 \; \frac{M_\odot}{\mathrm{kpc}^3} $.

During collisionless gravitational dynamics, phase mixing
increases the density and velocity dispersions in such a way
that the coarse grained phase space density either remains constant
or {\bf diminishes},  namely
\be
\frac{\rho}{\sigma^3} \leq 6.611 \times
10^8  \;  \mathcal{D} \;  \Big[\frac{m}{\mathrm{keV}}\Big]^4  \;
\frac{M_\odot/\mathrm{kpc}^3}{\big(\mathrm{km}/\mathrm{s}\big)^3}  \;
\,. \label{Dcon}
\ee
where $ \mathcal{D} $ is given by eq.(\ref{Dex})
for an arbitrary distribution function. For a particle that
decouples when it is ultrarelativistic $ \mathcal{D} $ does not depend on
the mass, hence eq.(\ref{Dcon}) yields a \emph{lower} bound on the mass
of the particle directly from the observed phase space density and
the knowledge of the distribution function.

\medskip

For comparison it is convenient to gather the values $ \mathcal{D} $ eq.(\ref{Dex})
for the usual LTE cases that follow from eqs.(\ref{dist}) and (\ref{MBf})
\be\label{LTDD}
\mathcal{D} = g \times \left\{  \begin{array}{l}
                                  1.963\times 10^{-3}~~\mathrm{Fermions},~\mu_d=0 \\
                                               3.657\times 10^{-3}~~\mathrm{Bosons~without~BEC} \\
                                                                    3.657\times 10^{-3}\,\Big(\frac{T_c}{T_d}
                                  \Big)^{ \frac{15}{2}}~~\mathrm{Bosons~with~BEC},~T_c>T_d  \\
                                                                    8.442 \times
                     10^{-2} \;  g_d \; Y_\infty~~\mathrm{non-relativistic~Maxwell-Boltzmann.}
                                \end{array} \right.
\ee
where $ g_d $ is the number of ultrarelativistic degrees of freedom at decoupling.

We note that for $ T_c \gg T_d $ the presence of a BEC increases
dramatically the primordial phase space density. This   is a
consequence of the {\it enhancement} of the particle density over the
thermal case due to the presence of the condensate, and the
\emph{decrease} in the velocity dispersion because the particles in
the condensate all have zero momentum.

\subsection{New bounds from phase space density and dShps-data}

We derive here new bounds from the latest compilation presented in
ref. \cite{gilmore} directly on $ \rho/\sigma^3 $ for the dataset
comprising ten satellite galaxies in the Milky-Way dSphs. It
proves convenient to write eq.(\ref{Dcon}) as
\be
m^4 \geq \frac{\Big[62.36~\mathrm{eV}\Big]^4}{\mathcal{D}} \; 10^{-4} \;
\frac{\rho}{\sigma^3} \; \frac{\big(\mathrm{km}/\mathrm{s}\big)^3}{M_\odot/\mathrm{kpc}^3}
\; , \label{equiv}
\ee
the data in ref. \cite{gilmore} yields the range
\be
0.9 \leq 10^{-4} \; \frac{\rho}{\sigma^3} \;
\frac{\big(\mathrm{km}/\mathrm{s}\big)^3}{M_\odot/\mathrm{kpc}^3}
\leq 20 \; ,\label{datars}
\ee
and we choose a fiducial value for this quantity in the middle of the range of the data
\cite{gilmore} $ \sim 5-10 $, leading to the new bound
\be \label{nboun}
m \gtrsim \frac{100}{\mathcal{D}^{\frac{1}{4}}} ~\mathrm{eV} \; .
\ee
For thermal relics that decoupled while ultrarelativistic with vanishing chemical potentials
and no BEC, we find from eqs.(\ref{LTDD}) and (\ref{nboun}),
\be  \label{m2bo}
m \gtrsim \frac1{g^\frac14} \Bigg\{\begin{array}{l}
                                            0.475~\mathrm{keV}~~\mathrm{Fermions} \\
0.407~\mathrm{keV}~~\mathrm{Bosons~without~BEC}
                                          \end{array}
\ee
and for Bosons with BEC ($T_d<T_c$) we find
\be \label{m2BEC}
m \gtrsim  \frac{1}{g^{\frac14}} \; 0.407~\mathrm{keV} \;
\Bigg[\frac{T_d}{T_c}\Bigg]^{\frac{15}{8}} \quad \mathrm{Bosons~with~BEC} \; .
\ee
For particles that decouple \emph{out of LTE} with arbitrary
distribution functions the form of the new bound is given by
eq.(\ref{nboun}) with $ \mathcal{D} $ given by eq.(\ref{Dex}). The
detailed form of $ \mathcal{D} $ is completely determined by the
distribution function at decoupling, which must be obtained from a
microscopic calculation of the kinetics of decoupling. Once the
distribution function is obtained, the new bound
eq.(\ref{nboun}) yields the {\bf lower bound} of the mass consistent with
the observational data.

Combining the \emph{upper} bound (\ref{bound}) with the \emph{lower}
bound eq.(\ref{equiv}) we establish the mass range for the DM candidate
\be    \label{massrange}
\frac{62.36~\mathrm{eV}}{\mathcal{D}^{\frac14}} \; \Bigg[10^{-4} \;
\frac{\rho}{\sigma^3} \; \frac{\big(\mathrm{km}/\mathrm{s}\big)^3}{M_\odot/\mathrm{kpc}^3}
\Bigg]^{\frac14} <  m \leq 2.695 \; \mathrm{eV} \; \frac{2  \; g_d \; \zeta(3)}{g \;
\int^\infty_0 y^2  \; f_{d }(y) \; dy} \; ,
\ee
where $ \mathcal{D} $ is given by eq.(\ref{Dex}) and the compilation
of  data in \cite{gilmore} constrains the bracket
$ \left[\cdots\right]^{\frac{1}{4}} \sim 1-2 $.

For thermal relics that decoupled in LTE while ultrarelativistic,
and taking the bracket in the middle of the range we obtain from
eqs.(\ref{LTDD}) and (\ref{massrange}) \bea   && \frac{444~
\mathrm{eV} }{g^{\frac14}}  \leq m \leq \frac{g_d}{g}~ 4.253 ~
\mathrm{eV}  ~~\mathrm{fermions~with}~ \mu_d=0  \; ,\nonumber \\
  && \frac{ 380~\mathrm{eV}}{g^{\frac14}}  \leq m \leq
\frac{g_d}{g}~ 2.695 ~\mathrm{eV}   ~~\mathrm{bosons~with} ~ \mu_d=0
\; \mathrm{and~no~BEC} \; \, \nonumber \\   &&
\frac{380~\mathrm{eV}}{g^{\frac14}}\Bigg[\frac{T_d}{T_c}\Bigg]^\frac{15}{8}
\leq m \leq \frac{g_d}{g}~ 2.695 ~\Bigg[\frac{T_d}{T_c}\Bigg]^3  \;
\mathrm{eV}  ~~\mathrm{BEC} \; . \label{massrangeF}\eea Therefore,
if the thermal relic decouples in \emph{equilibrium} this mass range
indicates that it \emph{must} decouple when $ g_d \; g^{-\frac34} >
110-150 $, namely at or above the electroweak scale \cite{kt}. In
the BEC case, for $ T_d\ll T_c $ the fulfillment of the bound
requires very large $ g_d \; g^{-\frac34} $, namely thermal
decoupling at a scale much larger than the electroweak scale.

An alternative is that the particle  is very weakly coupled to the
plasma and decouples \emph{away} from equilibrium with a
distribution function that yields a \emph{smaller abundance}
increasing the right hand side of eqn. (\ref{massrangeF} ).

\subsection{Generalized Tremaine-Gunn bound.}

The Tremaine-Gunn bound \cite{TG} establishes a relation between the
properties of dark matter in galaxies through their phase space
densities. It assumes that dark matter could be reliably described
by an isothermal sphere solution of the Lane-Emden equation with the
equation of state (\ref{isoeqnofstate}) \cite{bt,gas}. In thermal
equilibrium the quantity \cite{gas} \be\label{L} \eta = \frac{G \;
m^2 \; N}{L \; T} = \frac{2 \, G \; \rho \; L^2}{3 \; \sigma^2} \ee
  is bound to be  $ \eta \lesssim 1.6 $ to prevent the
gravitational collapse of the gas. Here $ V = L^3 $ stands for the
volume occupied by the gas,  $ N $ for the number of particles and $
T = \frac32 \; m \; \sigma^2 $ for the gas temperature. The length $
L $ is similar to the King radius \cite{bt}. However, the King
radius follows from the singular isothermal sphere solution while $
L $ is the characteristic size of a stable isothermal sphere
solution \cite{gas}.

Combining eq.(\ref{L})  with eq.(\ref{Dcon})
results in a \emph{generalized} Tremaine-Gunn bound
\be
m^4 \geq  \frac{\eta}{2 \; \sqrt3 \; G \;  L^2  \; \sigma \; \mathcal{D}}
= \eta \; \frac{\big[85.22 \;
\mathrm{eV}\big]^4}{\mathcal{D}} \; \frac{10\,\mathrm{km}/\mathrm{s}}{\sigma} \;
\Bigg[\frac{\mathrm{kpc}}{L} \Bigg]^2 \label{TGG}
\ee
therefore the
\emph{generalized} Tremaine-Gunn bound on the mass becomes
\be \label{GTG}
m \geq \frac{85.22 \; \mathrm{eV}}{\mathcal{D}^{\frac14}} \; \eta^\frac14 \;
\Bigg[\frac{10\,\mathrm{km}/\mathrm{s}}{\sigma}\Bigg]^{\frac14}
\Bigg[\frac{\mathrm{kpc}}{L} \Bigg]^\frac12 \; .
\ee
The compilation of recent photometric and kinematic data from ten
Milky Way dSphs satellites \cite{gilmore} yield values for the one
dimensional velocity dispersion $(\sigma)$  and the  radius
($ L $)  in the ranges
\be
0.5~\mathrm{kpc} \leq L \leq 1.8
~\mathrm{kpc} \quad , \quad
6.6~\mathrm{km}/\mathrm{s} \leq \sigma \leq 11.1
\mathrm{km}/\mathrm{s}\,. \label{gil}
\ee
 For particles that decouple in LTE when they are ultrarelativistic
(ultrarelativistic thermal relics)
 with vanishing chemical potential and no BEC we find from eqs.(\ref{LTDD}) and (\ref{GTG}),
\be \label{TRTG}
m \geq \left(\frac{\eta}{g}\right)^\frac14 \; \Bigg[
\frac{10\,\mathrm{km}/\mathrm{s}}{\sigma}\Bigg]^\frac14
\Bigg[\frac{\mathrm{kpc}}{L} \Bigg]^\frac12 \Bigg\{
\begin{array}{l}
                           0.405 ~ \mathrm{keV} ~~\mathrm{Fermions} \\
                          0.347 ~ \mathrm{keV} ~~\mathrm{Bosons}\; .
                 \end{array}
\ee
For the case of ultrarelativistic bosonic thermal relics with a BEC and $ T_d< T_c $ we find the bound
\be
m \geq  \left(\frac{\eta}{g}\right)^\frac14 \; 0.347 ~
\mathrm{keV} \; \Bigg[\frac{10\,\mathrm{km}/\mathrm{s}}{\sigma}\Bigg]^{\frac14}
\Bigg[\frac{\mathrm{kpc}}{L} \Bigg]^\frac12
\Bigg[\frac{T_d}{T_c}\Bigg]^\frac{15}8 \; . \label{TGBEC}
\ee
Therefore, the BEC case allows for {\it smaller masses} to saturate the
Tremaine-Gunn bound for $ T_c  \gg T_d $, a consequence of the
{\it enhanced} primordial phase space density in the presence of the BEC.

\subsection{DM mass values from velocity dispersion}

We can use the independent data provided in ref. \cite{gilmore} on
the mean density and velocity dispersion to explore bounds solely
from the velocity dispersion. Since the phase space density only
\emph{diminishes or remains constant} during the collisionless
gravitational dynamics of clustering, from which it follows that
\be\label{const} \frac{\rho_{DM}}{\sigma^3_{DM}} \geq
\frac{\rho_{s}}{\sigma^3_{s}} \; , \ee where $ \rho_{DM} $ and $
\sigma_{DM} $ are, respectively, the matter density and velocity
dispersion of the \emph{homogeneous} dark matter prior to gravitational
collapse.  $ \rho_s $ and $ \sigma_s $ are, respectively,
the satellite's mean volume mass density and velocity dispersion.
Assuming that DM has a single component, its density today
is \cite{pdg}
\be \label{dmhoy}
\rho_{DM} = \Omega_{DM} \; h^2 \;
1.054 \times 10^4  \; \mathrm{eV}/\mathrm{cm}^3  = 1.107\times 10^3
\; \mathrm{eV}/\mathrm{cm}^3   \;  ,
\ee
$ \sigma_{DM} $ is given by
eq.(\ref{sigdmz}). Ref. \cite{gilmore} quotes the following values for the favored
satellite's cored dark matter density and velocity dispersion
\be
\label{dat} \rho_s \sim 5 \; \frac{\mathrm{GeV}}{\mathrm{cm}^3}
\quad , \quad \sigma_s \sim 10 \; \frac{\mathrm{km}}{s} \; .
\ee
Eqs. (\ref{const}), (\ref{dmhoy}) and (\ref{dat}) lead to
\be\label{dms}
\sigma_{DM} \leq 0.06 \; \frac{\mathrm{km}}{s} \; .
\ee
Combining eq.(\ref{sigdmz}) for $ z = 0 $ and eq.(\ref{dms}) yields
\be\label{masdisp}
\frac{m}{\mathrm{keV}} \geq  \frac{0.847}{g^{\frac13}_d} \;
\Bigg[ \frac{\int_0^\infty y^4  \; f_d(y) \;
dy}{\int_0^\infty y^2 \;  f_d(y) \; dy}\Bigg]^{\frac12} \; .
\ee
For thermal fermions or bosons without chemical potential (no BEC) and $ 10 < g_d \lesssim 100 $ we
find $ m \sim 0.6-1.5~\mathrm{keV} $ in   agreement with the bounds found above and the
conclusions of ref. \cite{faber}. A suppression factor $ (T_d/T_c)^3 $ appears in the
BEC case for the same range of $ g_d $.

We emphasize that the bound eq.(\ref{masdisp}) is  \emph{independent}
from the bound eq.(\ref{m2bo}) obtained from the phase space density above,
and relies on the fact that the
observational data \cite{gilmore} yields separate information on $ \rho_s $ and
$ \sigma_s $.

It proves illuminating to analyze the velocity dispersion $ \sigma_{DM} $
from expression (\ref{sigdmz}) at $ z=0 $ for thermal relics. We find
\be \label{sigsdm}
\sigma_{DM} = \frac{1 }{g^\frac13_d}
\; \frac{\mathrm{keV}}{m}  \; \frac{\mathrm{km}}{\mathrm{s}}
  \left\{  \begin{array}{l}
                               0.187~\mathrm{Fermions}~\mu_d=0 \\
                                        0.167~\mathrm{Bosons~no~BEC} \\
                                0.167\times\big(\frac{T_d}{T_c}\big)^\frac{3}{2}~
                                       \mathrm{Bosons~with~BEC},~T_c>T_d\\
                         0.09\times \sqrt{x_d}~\mathrm{non-relativistic}
                                                        \end{array}    \right.
\ee
We see that for $ T_c\gg T_d $ light Bosonic particles that decoupled while
ultrarelativistic but undergo BEC can effectively act as CDM with {\it very small}
velocity dispersion.

\medskip

In ref. \cite{dominik} it is found that kinetic decoupling for a
WIMP of mass $ m \sim 100~\mathrm{GeV} $ occurs at $ T_d \sim 10
~\mathrm{MeV} $, leading to the estimate $ \sqrt{x_d} = \sqrt{m/T_d}
\sim 100 $. Thus, for CDM from weakly interacting massive particles
the velocity dispersion eq.(\ref{sigsdm}) is: \be \label{sigcdm}
\sigma_{wimp} \sim 10^{-8}  \; \Big( \frac{100 \;
\mathrm{GeV}}{m}\Big) \; \frac{\sqrt{x_d}}{100} \; g^{-\frac13}_d \;
\Big(9  \; \frac{\mathrm{km}}{\mathrm{s}}  \Big)  \; . \ee Thus, $
\sigma_{wimp} $ is {\bf eight orders of magnitude smaller} than the
typical velocity dispersion in dSphs
 \cite{gilmore} for wimps of $ m\sim 100 $ GeV that decoupled in LTE at $T_d
\sim 10\,\mathrm{MeV}$ \cite{dominik}.

\medskip

It is noteworthy to compare the phase space densities of the
homogeneous dark matter distribution for the thermal relics that
decoupled ultrarelativistically and non-relativistically with that
observed in the satellites dShps. If the distribution of dark matter
is cored \cite{gilmore}\footnote{$(\mathrm{eV}/c^2)/\mathrm{cm}^3 =
0.026\, M_\odot/\mathrm{kpc}^3$} \be \label{rss} \left(
\frac{\rho_s}{\sigma^3_s}\right)_{cored} \sim 5\times 10^6
~\frac{\mathrm{eV}/\mathrm{cm}^3} {\Big( \mathrm{km}/\mathrm{s}
\Big)^3} \quad  \; . \ee If the distribution of dark matter is
cusped, ref. \cite{gilmore} gives the value for the density $ \rho_s
\sim 2 \; \mathrm{TeV}/\mathrm{cm}^3 $ yielding \be\label{cusp}
\left(\frac{\rho_s}{\sigma^3_s}\right)_{cusped} \sim 2\times 10^9
~\frac{ \mathrm{eV}/\mathrm{cm}^3} {\Big( \mathrm{km}/\mathrm{s}
\Big)^3}  \quad   \; . \ee Assuming that a thermal relic  that
decoupled when ultrarelativistic is the \textbf{only} DM component
with the density given by the value today  $ \rho_{DM} \sim
1.107\times 10^3 \mathrm{eV}/\mathrm{cm}^3 $ \cite{pdg}, we find
from eqs.(\ref{sigdmz}) at $ z = 0 $, \be \label{tr}
\frac{\rho_{DM}}{\sigma^3_{DM}} \sim 10^6 ~\frac{
\mathrm{eV}/\mathrm{cm}^3} {\Big( \mathrm{km}/\mathrm{s}
\Big)^3}~\Bigg( \frac{m}{\mathrm{keV}}\Bigg)^3  \; g_d \;
\Bigg\{\begin{array}{l}
         0.177~~~\mathrm{Fermions} \\
              0.247~~~\mathrm{Bosons ~ without ~ BEC} \\
0.247\,(T_c/T_d)^\frac92 ~~~\mathrm{Bosons ~ with  ~ BEC}
     \; .
          \end{array}
\ee
Thus, for $ g_d > 10 $ we see that for $ m \sim \mathrm{keV} $
the phase space density for thermal relics that decoupled being
ultrarelativistic is of the \emph{same order} as the phase space
density in dShps with cores, eq.(\ref{rss}). Thermal
relics with mass in the $ \sim \mathrm{keV} $ range  obviously favor
{\it cores} over cusps because the primordial phase space is $
\rho_{DM}/\sigma^3_{DM} \geq \rho_s/\sigma^3_s $ for cores while $
\rho_{DM}/\sigma^3_{DM} \ll \rho_s/\sigma^3_s $ for a cuspy
distribution, and according to the theorem in \cite{theo,TG}, the
phase space density can only diminish during gravitational
clustering.

An enhancement factor $ (T_c/T_d)^\frac92 $ appears in eq.(\ref{tr})
for the case of a BEC. Notice, that for $ T_c/T_d \gtrsim 10 $ and
$ m\sim \mathrm{keV} $, a BEC \emph{yields a phase space density
consistent with cusps} as a result of the \emph{small} velocity dispersion
and the CDM behavior.

\medskip

Recent $N$-body simulations \cite{numQ} indicate that the phase
space density decreases by a factor $ 10-10^2 $ due to gravitational
relaxation during structure formation between $ 0\leq z\leq 10 $,
with smaller relaxation in WDM than in CDM \cite{numQ,dalcanton1}.
Therefore, from these numerical results it follows  that \be
\label{rela2} \frac{\rho_s}{\sigma^3_s} \sim 10^{-2} \;
\frac{\rho_{DM}}{\sigma^3_{DM}} \quad  \; . \ee Combining this
result with the observational results eqs.(\ref{rss})-(\ref{cusp})
and the primordial phase space density eq.(\ref{tr}) for a
\emph{thermal relic} that decoupled while ultrarelativistic, we find
\be \label{caropico} m_{cored} \sim \frac{15}{g^\frac13_d} \;
\mathrm{keV} \quad , \quad m_{cusp} \sim \frac{100}{g^\frac13_d} \;
\mathrm{keV} \; . \ee These values and the {\bf upper} bounds for $
m $ in eqs.(\ref{massrangeF}) yield the following bounds  for
\emph{thermal relics} \be \label{gcapi} g_d \; g^{-\frac34} \geq 500
\quad {\rm for ~ cores} \quad , \quad g_d \; g^{-\frac34} \geq 2000
\quad {\rm for ~ cusps} \; . \ee Therefore, \emph{thermal relics},
DM candidates that decouple when relativistic, must decouple at a
temperature well {\it above} the electroweak scale.
Eqs.(\ref{caropico}) and (\ref{gcapi}) imply for the mass value: \be
\label{masarango} m_{cored}\sim \frac2{g^\frac14} \; \mathrm{keV}
\quad , \quad m_{cusp} \sim\frac8{g^\frac14} \; \mathrm{keV} \; .
\ee Although $ m_{cusp} $ is not too much larger than $ m_{cored} $
it is noteworthy that the thermal relic DM candidate that leads to
cusped profiles must decouple when $ g_d  \gtrsim 2000 $ namely very
early at a temperature scale corresponding to a grand unified theory
with a large symmetry  group.

\medskip

For the case of CDM from wimps which decoupled non-relativistic, we
find from eqs.(\ref{dmhoy}) and (\ref{sigsdm}) \be\label{denw}
\frac{\rho_{wimp}}{\sigma^3_{wimp}} \sim 10^{24} \; \frac{
\mathrm{eV}/\mathrm{cm}^3}{\Big( \mathrm{km}/\mathrm{s}
\Big)^3}~\Bigg( \frac{m}{100\,\mathrm{GeV}}\Bigg)^3 \;
\left(\frac{100}{\sqrt{x_d}}\right)^3 \; g_d  \; . \ee The  phase
space density always decreases by dynamical relaxation, a result
recently confirmed numerically by $N$-body simulations\cite{numQ}.
For initial values of the phase space density which are much lower
than the primordial ones, these yield a typical decrease  by a
factor $ 10^2 - 10^3 $ \cite{numQ}. If these results should persist
in N-body simulations with larger values of the initial phase space
density, they would imply a tension between  the phase space density
of WIMPs eq.(\ref{denw}) being eighteen to fifteen orders of
magnitude larger than that in dShps either cored eq.(\ref{rss}) or
cusped eq.(\ref{cusp})\cite{gilmore}.

Combining eqs.(\ref{rss}), (\ref{cusp}), (\ref{rela2}) and (\ref{denw}) yield for wimps as DM,
\be\label{final}
\sqrt{m \; T_d} \sim \frac{10}{g^\frac13_d} \; \mathrm{keV} \quad {\rm for ~ cores}
\quad , \quad
 \sqrt{m \; T_d} \sim \frac{100}{g^\frac13_d} \; \mathrm{keV}\quad {\rm for ~ cusps} \; .
\ee

>From the combined  analysis of the primordial phase space densities, the
observational data on dSphs \cite{gilmore} and the $N$-body
simulations in ref. \cite{numQ}, we conclude the following:
\begin{itemize}
\item{\textbf{(i):}  Thermal relics with $ m \sim $ few keV that
decouple when ultrarelativistic lead to a primordial phase space density
of the same order of magnitude as in cored dShps and
disfavor cusped satellites for which the data \cite{gilmore}
yields a much larger phase space density. }

\item{\textbf{(ii):}
Light bosonic particles decoupled while ultrarelativistic and which form a BEC
lead to phase space densities consistent with cores and if $ T_c/T_d \gtrsim 10 $,
also consistent with cusps. However, for thermal relics to satisfy the
bound eq.(\ref{massrangeF}) they must decouple when $ g_d \; g^{-\frac34} \; (T_d/T_c)^\frac98
> 130 $, namely above the electroweak scale. Recall that typically $ g $ takes a value between
one and four.}
\end{itemize}

\section{Non-equilibrium effects:}

The main results of our analysis are the new bounds from DM
abundance and phase space density of dShps summarized in
eq.(\ref{massrange}). When the dark matter candidate decouples
\emph{out of LTE}   these bounds establish a direct connection with
the {\it microphysics} via the frozen distribution functions. These
functions must be obtained from a detailed calculation of the
{\it microscopic processes} that describe the production and pathway
towards equilibration of the corresponding dark matter candidate. If
kinetic (and chemical) freeze out occur out of LTE the distribution
functions will keep memory of the initial state and the detail of
the processes that established it.

\medskip

Non-equilibrium effects have been mainly considered for massive
particles that decoupled when non-relativistic \cite{steenneq} or as
distortions in the neutrino distribution functions during BBN
\cite{manga,kawa}. Instead, we focus here
  on DM constraints from  decoupling out of (LTE) at temperatures larger
  than the BBN scale and  when particles are ultrarelativistic.
  {\it Decoupling out of LTE} in this case has been much less studied. In
  this section we explore a cosmologically relevant mechanism of production
  and equilibration which describes a wide variety of situations out of LTE.

\subsection{Particle production followed by an UV cascade:}

Early studies of particle production via parametric amplification and
oscillations of inflaton-like scalar fields revealed that particles
are produced via this mechanism primarily in a low momentum band of
wave vectors \cite{boydata} leading to a non-thermal spectrum
(figs. 2-3 in ref. \cite{boydata} illustrate these effects).

Subsequent studies \cite{dvd}
showed that the early phase of parametric amplification and particle
production is followed by a long stage of mode mixing and scattering
that redistributes the particles: the larger momentum modes are populated
by a \emph{cascade} whose front moves towards the ultraviolet
akin to a direct cascade in turbulence, leaving in its wake a state
of nearly LTE but with a {\it lower} temperature than that of
equilibrium \cite{dvd}.

The dynamics during the cascade process diminishes the amplitude of
the distribution function at lower momenta and populates the higher
momentum modes. The distribution function develops a \emph{front}
that moves towards the ultraviolet. Behind the front the
distribution function is nearly that of LTE with a different
temperature and amplitude  and slowly evolves towards thermal
equilibrium \cite{dvd}. If these particles are very weakly coupled
to the plasma it is possible that the advance of the cascade and the
front of the distribution towards larger momenta is
\emph{interrupted} when the rate of scattering or mode mixing
becomes smaller than the expansion rate. In this case, the
distribution function is \emph{frozen} well before reaching complete
LTE resulting in a population of modes primarily at lower momenta up
to the scale of the front. This study \cite{dvd} suggests the
following frozen distribution function
\be
f_d(y) = f_0 \; f_{eq}\Big(\frac{y}{\xi}\Big) \; \theta(y_0-y) \label{cascade} \; ,
\ee
where $ f_{eq}\big(\frac{p_c}{\xi T_d}\big) $ is the
equilibrium distribution function for an ultrarelativistic particle
at an effective temperature $ \xi \; T_d $. Namely, $ \xi = 1 $ at thermal equilibrium
and $ \xi < 1 $ {\bf before}  thermodynamical equilibrium is attained.

This form describes fairly accurately the cascade with a \emph{front} that moves
towards the ultraviolet, which is \emph{interrupted} at a fixed
value of the momentum, identified here to be $ p ^0_c= y_0  \; T_d $;
 $ T_d $ is the temperature of the environmental degrees of
freedom that are in LTE at the time of decoupling.

The amplitude $ f_0 $ and effective temperature $ \xi  \; T_d \leq T_d $ reflect an
incomplete thermalization behind the front of the cascade and
determine the average number of particles in its \emph{wake}
\cite{dvd}. This interpretation is borne out by the detailed
numerical studies in ref. \cite{dvd}. For Fermi-Dirac
ultrarelativistic particles (with vanishing chemical potential) $
0\leq f_0 \leq 2 $ whereas for Bose-Einstein ultrarelativistic
particles $ 0 \leq f_0 \leq \infty $. Neglecting the possibility of
a BEC, for a fermionic or bosonic equilibrium distribution
function  $ f_{eq} $, we find
\bea\label{bigF}
&& \int^\infty_0 y^2
\; f_d(y) \;  dy = f_0  \; \xi^3  \; F\Big[ \frac{p^0_c}{\xi \;
T_d}\Big] \quad , \quad F(s) = \int^s_0 y^2  \; f_{eq}(y) \; dy  \;
, \cr \cr && \int^\infty_0 y^4 \; f_d(y) \;  dy = f_0  \; \xi^5  \;
G\Big[ \frac{p^0_c}{\xi\; T_d}\Big] \quad , \quad G(s) = \int^s_0
y^4  \; f_{eq}(y) \; dy  \; .
\eea
and the primordial phase space density becomes
\be \label{ache}
\mathcal{D} = f_0  \;
\mathcal{D}_{eq} \; H(s)  \quad , \quad s=\frac{y_0}{\xi}=
\frac{p^0_c}{\xi \; T_d}\quad , \quad H(s) \equiv
\left[\frac{F(s)}{F(\infty)}\right]^\frac52 \;
\left[\frac{G(\infty)}{G(s)}\right]^\frac32 \; ,
\ee where $
\mathcal{D}_{eq} $ is the phase space density eq.(\ref{Dex}) for the
equilibrium distribution $ f_{eq} $.

\medskip

\begin{figure}[h]
\begin{center}
\includegraphics[height=2.5in,width=2.2in,keepaspectratio=true]{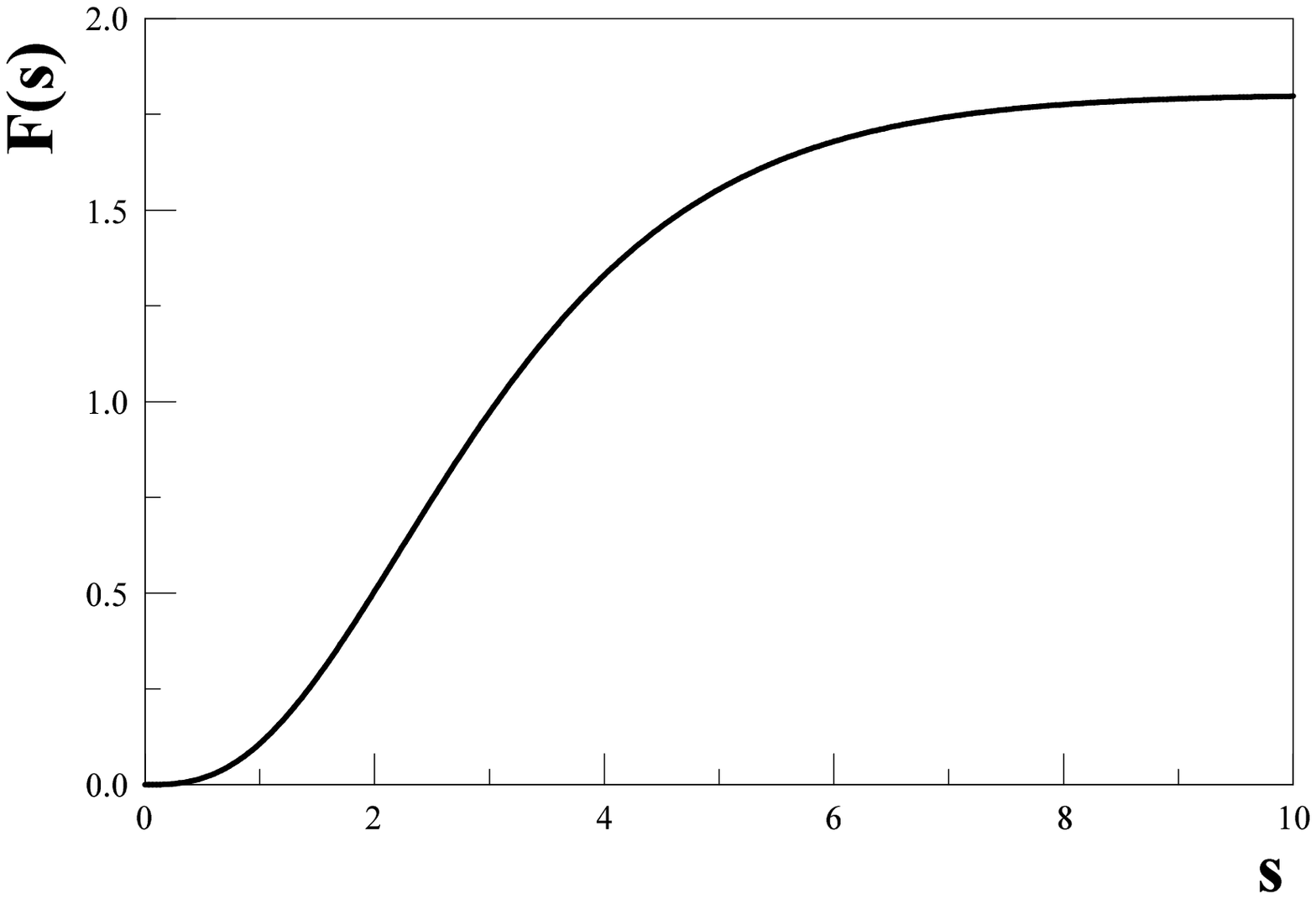}
\includegraphics[height=2.5in,width=2.2in,keepaspectratio=true]{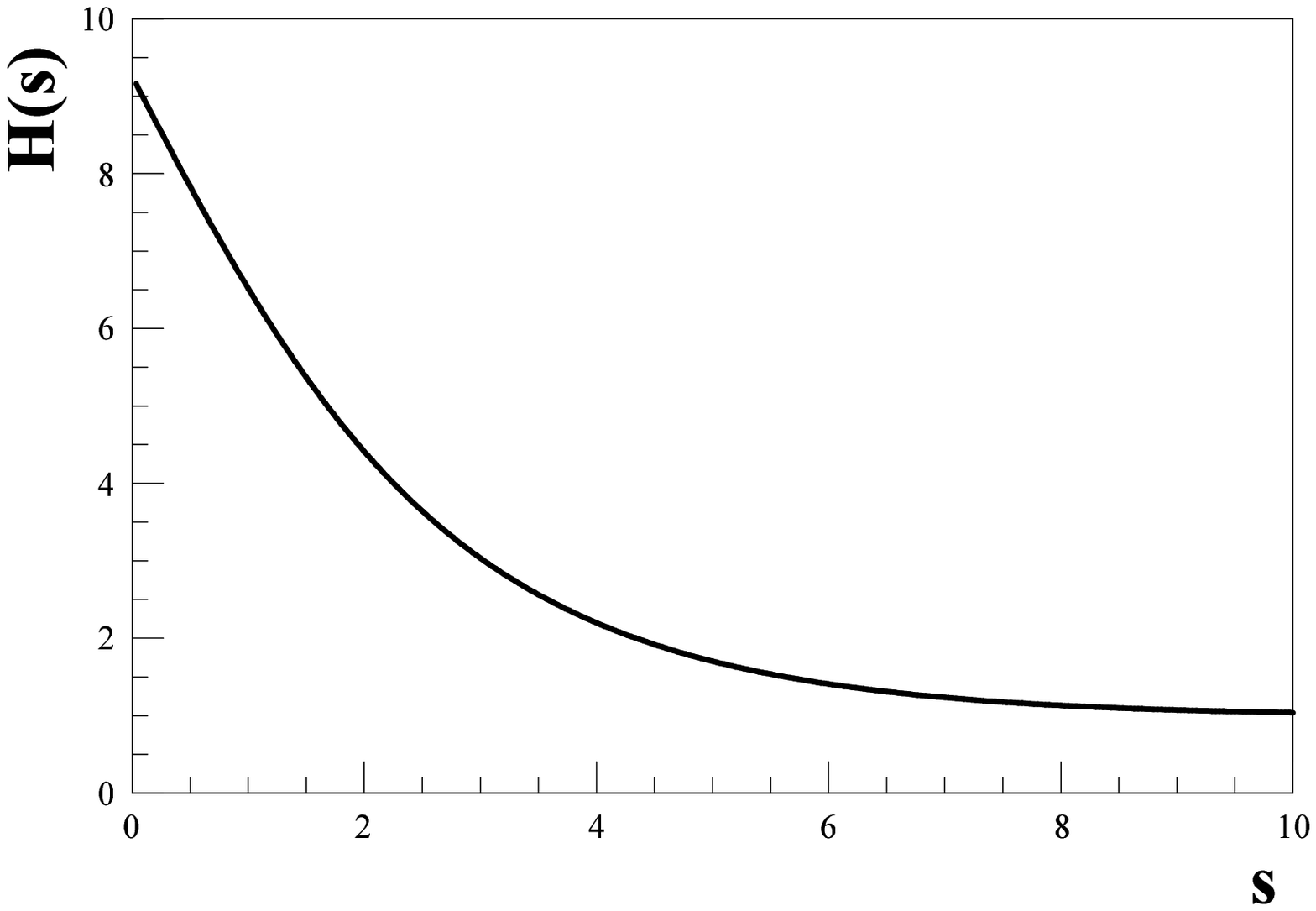}
\includegraphics[height=2.5in,width=2.2in,keepaspectratio=true]{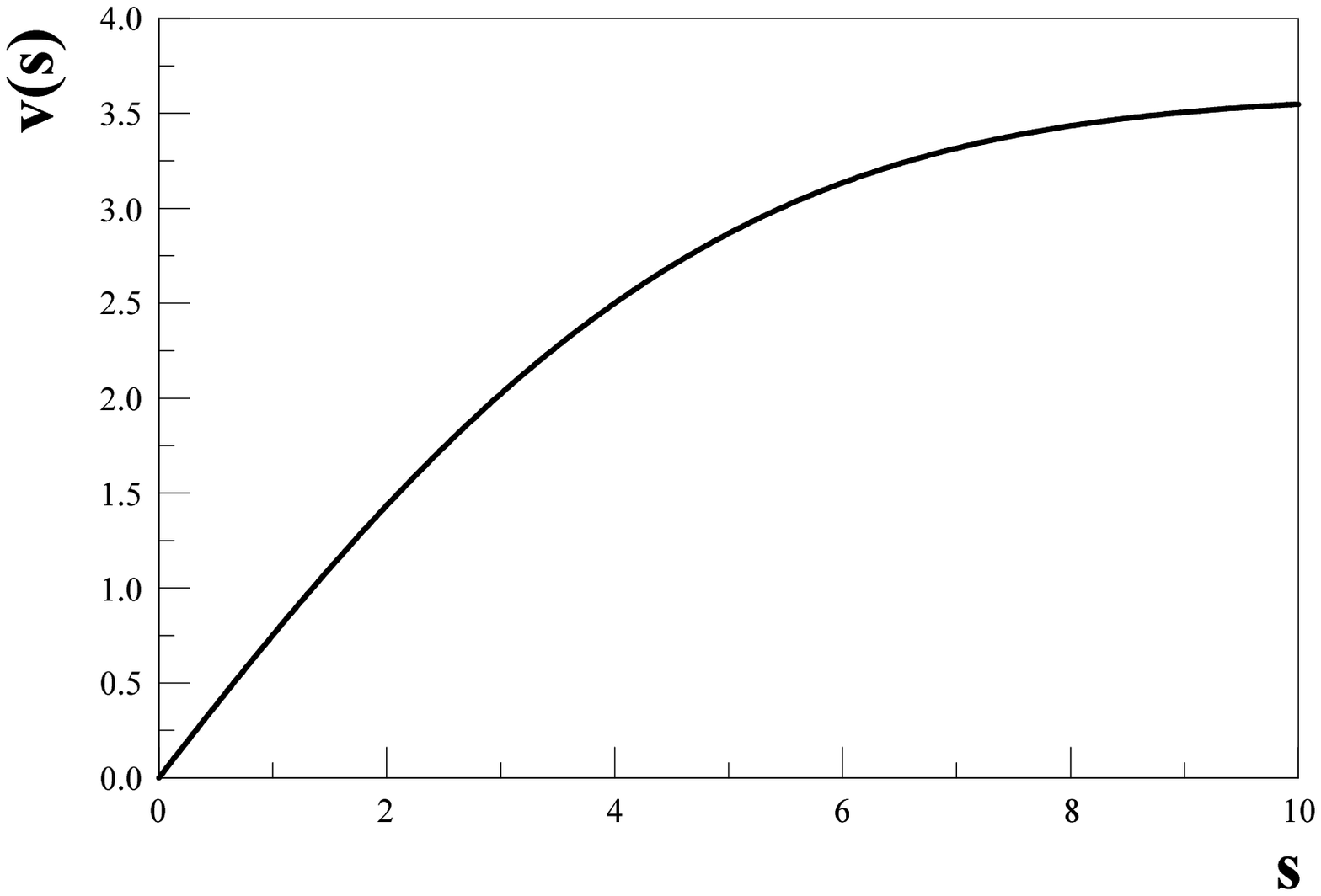}
\caption{The functions $ F(s) $ (left panel), $ H(s) $ (middle panel)
and $ v(s) $ (right panel) vs. $ s=y_0/\xi $,  for $ f_{eq} $   the
Fermi-Dirac distribution function without chemical potential. }
\label{fig:nofeq}
\end{center}
\end{figure}

For a fermionic species without chemical potential [$
f_{eq}(y)=1/(e^y+1) $], the bound eq.(\ref{massrange}) becomes \be
\frac{ 475 \; \mathrm{eV}}{\big[f_0  \; g  \; H(s) \big]^{\frac14}}
\leq m \leq \frac{g_d}{g}~ \frac{4.253}{f_0  \; \xi^3  \; F(s)}
~\mathrm{eV} \; , \label{massrangeeta} \ee and the one dimensional
velocity dispersion eq.(\ref{sigdmz}) becomes today: \be
\label{veldisnoneq} \sigma_{DM} = \frac{0.05124~\xi}{g^{\frac13}_d}
\; v(s) \; \Big(\frac{\mathrm{keV}}{m} \Big) \; \Big(
\frac{\mathrm{km}}{s} \Big) \quad , \quad v(s) \equiv
\sqrt{\frac{G(s)}{F(s)}} \; . \ee The functions $ F(s), \; H(s) $
and $ v(s) $ for the case $ f_{eq}(y) = 1/(e^y+1) $ are displayed in
fig. \ref{fig:nofeq}. For the Bose-Einstein case without a BEC the
behaviors of $ F(s), \; H(s) $ and $ v(s) $ are qualitatively
similar.

\medskip

It is clear that the bound eq.(\ref{massrangeeta}) for the range of
$m$ can easily be satisfied for moderate values $ g_d \sim 10-50 $
corresponding to decoupling temperatures $ 1\,\mathrm{MeV} \lesssim
T_d \lesssim 1\,\mathrm{GeV} $ and  $ f_0  \; \xi^3  \; F(s) \lesssim 0.08 $.

\medskip

Remarkably, the non-equilibrium distribution eq.(\ref{cascade})
turns out to be a generalization of several  non-equilibrium
distribution functions of cosmological relevance proposed in the
literature:

\begin{itemize}

\item{\textbf{(a):}
sterile neutrinos produced non-resonantly via the Dodelson-Widrow
mechanism \cite{dw} for which the distribution function is obtained
from (\ref{cascade}) by taking $ \xi =1;  \; s \rightarrow \infty,
\; f_0 \sim 0.043  \; \mathrm{keV}/m $ (ref. \cite{dw}). In this case
we find for the mass range, phase space density and velocity
dispersion respectively,
\be
\frac{1.04  \; \mathrm{keV}}{g^\frac14}
\leq m \leq \frac{g_d}{g} \;  56.5 \; \mathrm{eV} \quad , \quad
\frac{\rho_{DM}}{\sigma_{DM}^3} =5.7  \; g \times 10^4 \;
\Big[\frac{m}{\mathrm{keV}}\Big]^3  \;
\frac{M_\odot/\mathrm{kpc}^3}{\big(\mathrm{km}/\mathrm{s}\big)^3}\quad
, \quad \sigma_{DM} =   \frac{0.187}{g^{\frac13}_d} \;
\Big(\frac{\mathrm{keV}}{m}\Big)  \; \Big(\frac{\mathrm{km}}{s}
\Big)\; . \label{dwmb}
\ee
The major  uncertainty is the evaluation
of $ g_d $. In the Dodelson-Widrow \cite{dw} scenario the sterile
 neutrino production peaks at $ T\sim 130 \; \mathrm{MeV} $, this temperature is
very near the region where the QCD phase
 transition occurs at which the effective number of ultrarelativistic degrees of freedom
changes dramatically.
 \emph{If} decoupling occurs at a temperature higher than the QCD critical temperature,
then $ g_d\sim 30 $ and the
 mass bound eq.(\ref{dwmb}) may be fulfilled, but for a
 lower decoupling temperature when $ g_d \lesssim 25-30 $
the mass bound may not be fulfilled.
If the mass bound
 is fulfilled, $ \rho_{DM}/\sigma^3_{DM} $ is compatible with cored dSphs \cite{gilmore}
[see eq.(\ref{rss})] but not with
 the cusped distributions, [see eq.(\ref{cusp})]. Combining the bound eq.(\ref{dwmb}), the
 observed phase space density eq.(\ref{rss}) \cite{gilmore} and the
$N$-body results of ref. \cite{numQ}
which yield phase space relaxation by a factor $ \sim 10^2$ we find that
 \be
m\sim \frac4{g^{\frac13}} \; \mathrm{keV} \; .
\ee}
\item{\textbf{(b):}   sterile neutrinos produced by a net-lepton
number driven resonant conversion studied by Shi and
Fuller \cite{este} for which the   distribution function is obtained
from eq.(\ref{cascade}) for $ \xi=1, \; s\sim 0.7, \;f_0 \sim 1 $, (see fig.
1 in the first reference in \cite{este}). We find
\be
\frac{ 289  \; \mathrm{eV}}{g^\frac14} \leq m \leq \frac{g_d}{g} \; 98.9  \;
\mathrm{eV} \quad , \quad
\frac{\rho_{DM}}{\sigma^3_{DM}} = 9.6 \; g \times
10^6 \; \Big[\frac{m}{\mathrm{keV}}\Big]^4 \;
\frac{M_\odot/\mathrm{kpc}^3}{\big(\mathrm{km}/\mathrm{s}\big)^3} \quad , \quad
\sigma_{DM} =   \frac{0.028}{g^{\frac13}_d} \; \frac{\mathrm{keV}}{m} \;
\frac{\mathrm{km}}{s}\; .\label{fsb}
\ee
Again, a source of  uncertainty is the evaluation of $ g_d $, because in the
resonant-mediated sterile neutrino production,
the maximum production rate is near the QCD temperature \cite{este}. However, it
 is clear that in this case the mass bound is \emph{less sensitive}
 to the uncertainty in $ g_d $ (a small value $ g_d \sim 10 $
fulfills the bound), although the MSW resonance occurs also near the QCD critical
temperature \cite{este}.
The velocity dispersion is small because the distribution is skewed towards small momenta.
Again, $ \rho_{DM}/\sigma^3_{DM} $ is consistent with {\it cored} dSphs [see eq.(\ref{rss})]
but not with {\it cusped} distributions [see eq.(\ref{cusp})]. A similar analysis as
 in the previous case combining the observational data, the results of
ref. \cite{numQ} and the
bound eq.(\ref{fsb}) yields
\be
m\sim \frac{0.8}{g^{\frac14}}\,\mathrm{keV}\,.
\ee}
\item{\textbf{(c):} Our distribution function eq.(\ref{cascade}) for
$ \xi = 1; \; s\rightarrow \infty $ and $ f_0=\beta $ yields the
distribution function proposed in ref. \cite{strigari}  to model WDM [eq.(8) in
 ref. \cite{strigari}]. We find
\be
\frac{475 \; \mathrm{eV}}{(\beta \; g)^\frac14} \leq m \leq \frac{g_d}{\beta  \;
g}\;  2.36 \; \mathrm{eV} \quad , \quad
\frac{\rho_{DM}}{\sigma^3_{DM}} = 1.33 \; \beta  \; g \times 10^6 \;
\Big[\frac{m}{\mathrm{keV}}\Big]^4 \;
\frac{M_\odot/\mathrm{kpc}^3}{\big(\mathrm{km}/\mathrm{s}\big)^3}\quad
, \quad \sigma_{DM} =   \frac{0.187}{g^{\frac13}_d} \;
\frac{\mathrm{keV}}{m} \; \frac{\mathrm{km}}{s}\; .
\ee
The parameter  $ \beta $ cannot be too small, although a small $ \beta $
increases the mass, it decreases the phase space density.  }
\end{itemize}

Although recent studies \cite{boyho} suggest that    the  description
of the production mechanism   of sterile neutrinos must be
reassessed with likely implications on  their distribution functions
after decoupling, the above estimates provide a \emph{guidance} to
the range of mass, primordial phase space density and velocity
dispersions for sterile neutrinos as possible WDM candidates.

\section{Conclusions}

We have obtained new constraints on light DM candidates that
decoupled while ultrarelativistic in or out of LTE in terms of their
distribution functions. The only assumption is that these distribution functions are
homogeneous  and isotropic. A Liouville invariant coarse grained
primordial phase space density is introduced that allows to combine
phase space density arguments with a recent compilation of
photometric and kinematic data on dSphs galaxies to yield
\emph{new constraints} on the mass, velocity dispersion and
phase space density of DM candidates. The new constraint on
the mass range is
\be
\frac{62.36~\mathrm{eV}}{\mathcal{D}^{\frac{1}{4}}} \;
\Bigg[10^{-4}  \; \frac{\rho}{\sigma^3} \; \frac{\big(\mathrm{km}/\mathrm{s}\big)^3}{M_\odot/\mathrm{kpc}^3}
\Bigg]^{\frac14} \leq   m \leq 2.695 \; \frac{2  \; g_d \; \zeta(3)}{g \;
\int^\infty_0 y^2  \; f_{d }(y) \; dy} \; \mathrm{eV}   \; ,
\ee
where the primordial phase space density is given by
\be
\mathcal{D} = \frac{g}{2 \; \pi^2} \frac{\Bigg[\int_0^\infty y^2  \; f_d(y)  \; dy
\Bigg]^{\frac52}}{\Bigg[\int_0^\infty y^4  \; f_d(y) \; dy \Bigg]^{\frac32}} \; ,
\ee
$ f_d(p_c/T_d) $ is the distribution
function at decoupling, $ g $ the number of internal degrees of
freedom of the particle, and $ \rho/\sigma^3 $ is the phase space
density obtained from observations. The \emph{upper bound} arises
from requesting that the DM candidate has a density $ \leq \rho_{DM} $
today, and the \emph{lower} bound arises from requesting that the
phase space density in halos $ \rho/\sigma^3 $ be \emph{smaller} than
or equal to the primordial phase space density of the collisionless non-relativistic
(today) DM component
$$
\rho_{DM}/\sigma^3_{DM}=3^\frac32  \; m^4  \; \mathcal{D} .
$$
We have studied the consequences of Bose-Einstein condensation of
light ultrarelativistic particles when chemical freeze out occurs well
before kinetic decoupling at $ T_d < T_c $ with $ T_c $ the critical
temperature below which a non-vanishing condensate fraction exists.
We find that the presence of the condensate hastens the onset of the
non-relativistic regime and that Bose-Einstein condensed particles
can effectively act as a CDM component {\it even} when they decoupled being
ultrarelativistic. The reason for this unusual behavior is that
the   particles in the condensate all have vanishing velocity dispersion.

For \emph{thermal relics} we find
\be
\mathcal{D} = g \times \left\{  \begin{array}{l}
                                  1.963\times 10^{-3}~~\mathrm{Fermions},~\mu_d=0 \\
                                   3.657\times 10^{-3}~~\mathrm{Bosons~no~BEC} \\
                                  3.657\times 10^{-3}\,\Big(\frac{T_c}{T_d}
                                  \Big)^{ \frac{15}{2}}~~\mathrm{Bosons~with~BEC},~T_c>T_d  \\
                                                                  8.442\times
                                  10^{-2}\, g_d\,Y_\infty~~\mathrm{non-relativistic~Maxwell-Boltzmann.}
                                \end{array} \right.\label{LTDD2}
\ee
The combination of data in ref. \cite{gilmore} from dSphs when
applied to \emph{light thermal relics}  yields the mass range
 \bea
&& \frac{444~
\mathrm{eV} }{g^{\frac14}}  \leq m \leq \frac{g_d}{g}~ 4.253 ~
\mathrm{eV}  ~~\mathrm{fermions~with}~ \mu_d=0  \; ,\nonumber \\
  && \frac{ 380~\mathrm{eV}}{g^{\frac14}}  \leq m \leq
\frac{g_d}{g}~ 2.695 ~\mathrm{eV}   ~~\mathrm{bosons~with} ~ \mu_d=0
\; \mathrm{and~no~BEC} \; \, \nonumber \\   &&
\frac{380~\mathrm{eV}}{g^{\frac14}}\Bigg[\frac{T_d}{T_c}\Bigg]^\frac{15}{8}
\leq m \leq \frac{g_d}{g}~ 2.695 ~\Bigg[\frac{T_d}{T_c}\Bigg]^3  \;
\mathrm{eV}  ~~\mathrm{BEC} \; . \label{massrangeF2}
\eea
with the implication that if these particles are suitable DM candidates, they
must decouple at high temperature when the effective number of
ultrarelativistic degrees of freedom is $ g_d > 100 $. Namely, in absence
of a BEC, thermal decoupling must occur above the electroweak scale.
In the BEC case, for $ T_d\ll T_c $, the fulfillment of
the bound requires very large $ g_d $. Namely, in the presence of a
BEC thermal decoupling occurs at a scale much larger than the electroweak scale
for $ T_d\ll T_c $.

Assuming that the DM particle is the only
component with the density $ \rho_{DM} $ today, we obtained an
independent bound from velocity dispersion which for the favored
cored profiles \cite{gilmore} yield the lower mass bound
\be
\frac{m}{\mathrm{keV}} \geq  \frac{0.855}{g^{\frac13}_d}\,
\Bigg[ \frac{\int_0^\infty y^4  \; f_d(y)
 \; dy}{\int_0^\infty y^2 \;  f_d(y) \; dy}\Bigg]^{\frac12} \label{masdisp2} \; .
\ee
For light thermal relics this bound implies that $ m  \gtrsim 0.6-1.5\,\mathrm{keV} $ with a
 suppression factor $ T_d/T_c $ in the BEC case.

For light thermal relics that decoupled while ultrarelativistic we find the primordial
phase space density
\be
\frac{\rho_{DM}}{\sigma^3_{DM}} \sim 10^6 ~\frac{
\mathrm{eV}/\mathrm{cm}^3} {\Big( \mathrm{km}/\mathrm{s}
\Big)^3}~\Bigg( \frac{m}{\mathrm{keV}}\Bigg)^3  \; g_d \;
\Bigg\{\begin{array}{l}
         0.177~~~\mathrm{Fermions} \\
              0.247~~~\mathrm{Bosons ~ without ~ BEC} \\
0.247\,(T_c/T_d)^\frac92 ~~~\mathrm{Bosons ~ with  ~ BEC}
     \; .
          \end{array}
\ee
An enhancement factor $ (T_c/T_d)^\frac92 $ appears in the r.h.s. in the presence of a BEC.

For wimps with kinetic decoupling temperature $ 10 $MeV \cite{dominik}, we find
\be\label{wimpo}
\frac{\rho_{wimp}}{\sigma^3_{wimp}} \sim 10^{24} \; \frac{
\mathrm{eV}/\mathrm{cm}^3}{\Big( \mathrm{km}/\mathrm{s}
\Big)^3}~\Bigg( \frac{m}{100\,\mathrm{GeV}}\Bigg)^3   \; g_d  \; .
\ee
The observational data compiled in ref. \cite{gilmore} assuming a
favored cored profile suggests \be
\left(\frac{\rho_s}{\sigma^3_s}\right)_{cored} \sim 5\times 10^6
~\frac{ \mathrm{eV}/\mathrm{cm}^3} {\Big(
\mathrm{km}/\mathrm{s}\Big)^3}\,. \label{core2}
\ee
If the distribution of dark matter is cusped, ref. \cite{gilmore} gives the
value for the density $ \rho_s \sim 2 \; \mathrm{TeV}/\mathrm{cm}^3 $ yielding
\be
\left(\frac{\rho_s}{\sigma^3_s}\right)_{cusped} \sim
2\times 10^9 ~\frac{ \mathrm{eV}/\mathrm{cm}^3} {\Big(
\mathrm{km}/\mathrm{s} \Big)^3} \,.\label{cusp2}
\ee
Therefore, for
$ g_d \gtrsim 10 $ the primordial phase space density for
\emph{thermal relics} with $ m \sim \mathrm{keV} $ favors a {\bf
cored distribution}.

Notice that a bosonic thermal relic
that features a BEC can behave as CDM with small velocity
dispersion and a primordial phase space density consistent with
cusped distributions if $ T_d \ll T_c $. However,  these BEC DM
candidates must decouple at a temperature scale {\it higher} than the
electroweak.

Recent results from $N$-body simulations suggests that the phase
space density relaxes by a factor $ \sim 10^2 $ during gravitational
clustering for $ 0 \leq z \leq 10 $ \cite{numQ}. Combining these
numerical results with the observational results on
dSphs \cite{gilmore} and the present DM density, we conclude that the
mass of \emph{thermal relics} that decoupled when ultrarelativistic
is
\be \label{masarango2}
m_{cored}\sim \frac2{g^\frac14} \;
\mathrm{keV} \quad , \quad m_{cusp} \sim\frac8{g^\frac14} \;
\mathrm{keV} \; .
\ee
The decoupling temperature for the DM
candidate that would favor cusped profiles must be near a grand
unified scale for a large symmetry group with $g_d \gtrsim 2000$
which effectively results in a colder relic today with a far smaller
velocity dispersion.

\medskip

The \emph{enormous} discrepancy between the primordial phase space
density for WIMPs of $ m\sim 100\,\mathrm{GeV}; \; T_d \sim 10\,
\mathrm{MeV} $, eq.(\ref{wimpo}) and the phase space densities in
dSphs, either cored (eq.\ref{core2} ) or cusped (eq.\ref{cusp2})
cannot   be explained by the two orders of magnitude of
gravitational relaxation of phase space densities found with recent
$N$-body simulations \cite{numQ}, although these initialize the
simulation with much smaller values of the primordial phase space
density.

\medskip

We have studied a scenario for decoupling out of equilibrium
motivated by previous studies of particle production and
thermalization via an UV cascade. The distribution function obtained
from previous studies \cite{dvd}, remarkably describes the
non-equilibrium distribution functions for sterile neutrinos
produced either resonantly \cite{este} or non-resonantly \cite{dw} as
well as a recently proposed model for halo structure \cite{strigari}.
Our bounds in terms of arbitrary distribution functions
lead to the following bounds on the mass, phase space density and
velocity dispersion of these light relics that decoupled out of LTE:

\begin{itemize}

\item{For sterile neutrinos produced non-resonantly via the Dodelson-Widrow
mechanism \cite{dw}   we find
\be \frac{1.04
\; \mathrm{keV}}{g^\frac14} \leq m \leq \frac{g_d}{g} \;  46.5 \,
\mathrm{eV} \quad , \quad
\frac{\rho_{DM}}{\sigma^3_{DM}} = 0.57 \; g \times
10^5 \; \Big[\frac{m}{\mathrm{keV}}\Big]^3 \;
\frac{M_\odot/\mathrm{kpc}^3}{\big(\mathrm{km}/\mathrm{s}\big)^3} \quad , \quad
\sigma_{DM} =   \frac{0.187
 }{g^{\frac13}_d} \;
\Big(\frac{\mathrm{keV}}{m}\Big) \; \Big(\frac{\mathrm{km}}{s}\Big) \; . \label{dw2}
\ee
 The upper and lower bound on the mass can only be compatible if the sterile
neutrino decouples with $g_d \gtrsim 20-30$.
 For $m \sim \mathrm{keV}$ the primordial
 phase space density is compatible with cored but not with cusped profiles in the dShps data
\cite{gilmore}.
Combining these bounds with the results from $N$-body simulations on
the relaxation of the phase space density \cite{numQ} and with the
observational constraint eq.(\ref{rss}) \cite{gilmore}, we obtain the value
\be
m\sim \frac{4}{g^{\frac13}} \;   \mathrm{keV}
\label{boundste2}
\ee
for the mass of sterile neutrinos produced
non-resonantly by the Dodelson-Widrow mechanism.  }
\item{For  sterile neutrinos produced by a net-lepton
number driven resonant conversion \cite{este} we find
\be
\frac{289\,\mathrm{eV}}{g^\frac14} \leq m \leq \frac{g_d}{g}\;
81.4\; \mathrm{eV} \quad , \quad \frac{\rho_{DM}}{\sigma^3_{DM}} =
9.6 \; g \times 10^6\,\Big[\frac{m}{\mathrm{keV}}\Big]^4 \;
\frac{M_\odot/\mathrm{kpc}^3}{\big(\mathrm{km}/\mathrm{s}\big)^3}
\quad , \quad \sigma_{DM} = \frac{0.028 }{g^{\frac13}_d} \;
\Big(\frac{\mathrm{keV}}{m}\Big) \; \Big(\frac{\mathrm{km}}{s}\Big)
\; . \label{sterQ2}
 \ee
The small velocity dispersion is a consequence of the
 distribution function being skewed towards small momentum. Again
 for $ m \sim \mathrm{keV} $, the primordial
 phase space density is compatible with cored but not cusped profiles in the dShps
 data \cite{gilmore}. For sterile neutrinos produced by resonant conversion,
a similar analysis as for the  previous case yields
\be
m\sim \frac{0.8}{g^\frac14} \; \mathrm{keV} \; . \label{boundeste}
 \ee }
\item{For the model  proposed in ref. \cite{strigari} we find
\be \frac{475 \; \mathrm{eV}}{(\beta \; g)^\frac14} \leq m \leq
\frac{g_d}{\beta g} \; 1.94 \; \mathrm{eV} \quad , \quad
\frac{\rho_{DM}}{\sigma^3_{DM}} = 1.33 \; \beta \; g \times 10^6 \;
\Big[\frac{m}{\mathrm{keV}}\Big]^4 \;
\frac{M_\odot/\mathrm{kpc}^3}{\big(\mathrm{km}/\mathrm{s}\big)^3}
\quad , \quad \sigma_{DM} =   \frac{0.187}{g^{\frac13}_d} \;
\Big(\frac{\mathrm{keV}}{m}\Big) \; \Big(\frac{\mathrm{km}}{s}
\Big)\; . \label{bounstri}\ee }
\end{itemize}
It is noteworthy that the $N$-body results of ref. \cite{numQ} which
yield phase space relaxation by a factor $ \sim 10^2 $ bring the
values of the primordial phase space density of the above cases
within the range consistent with the phase space densities for cored
profiles in dSphs \cite{gilmore} for $ m \sim \mathrm{keV} $. On the
contrary, in the   case of WIMPs with $m\sim
100\,\mathrm{GeV},T_d\sim 10\,\mathrm{MeV}$, relaxation by {\bf
many} orders of magnitude is necessary for their phase space
densities to be compatible with the observed values both for cores
and for cusps.

Therefore the  bounds  eqs.(\ref{boundste2})-(\ref{boundeste})
confirm that $ \sim \mathrm{keV} $ relics that decouple out of
equilibrium while ultrarelativistic via the mechanisms described
above yield values for phase space densities that are in agreement
with cores in the DM distribution.

\medskip

The results obtained in this article for the new mass bounds,
primordial phase space densities and velocity dispersion in term of
arbitrary, but homogeneous and isotropic distribution functions
establish a link between the microphysics of decoupling and
observable quantities. They also warrant   deeper scrutiny  of the
non-equilibrium aspects of sterile neutrinos\cite{boyho} for a
firmer assessment of their potential as DM candidates.

\acknowledgements{We thank  Carlos Frenk,
for useful discussions, D.B. thanks  Andrew Zentner for
  fruitful discussions, and acknowledges support
 from the U.S. National Science Foundation through grant award
 PHY-0553418.}

\end{document}